\documentclass[english]{iopart_mod}

\usepackage[reqno,centertags,intlimits,fleqn]{amsmath}
\usepackage{amsfonts}
\usepackage{amssymb}
\usepackage{babel}
\usepackage{cite}
\usepackage{graphicx}

\usepackage{GEFI}

\begin{document}

\topical[Asymptotic directional structure of radiative fields]
{Asymptotic directional structure of radiative fields in
spacetimes with a cosmological constant}

\author{Pavel Krtou\v{s} and Ji\v{r}\'{\i} Podolsk\'y}

\address{
  Institute of Theoretical Physics, 
  Charles University in Prague,\\
  V Hole\v{s}ovi\v{c}k\'{a}ch 2, 180 00 Prague 8, Czech Republic
  }
\eads{\mailto{Pavel.Krtous@mff.cuni.cz}, \mailto{Jiri.Podolsky@mff.cuni.cz}}

\date{2.07; \today}

\begin{abstract}
We analyze the directional properties of general gravitational, electromagnetic, and
spin-$s\,$ fields  near conformal infinity $\scri$. The fields are evaluated in
normalized tetrads which are parallelly propagated along null
geodesics which approach a point $P$ of $\,\scri$. The standard peeling-off property
is recovered and its  meaning is discussed and refined. When the (local) character of the conformal infinity
is null, such as in asymptotically flat spacetimes, the dominant term which is
identified with radiation is unique. However, for spacetimes with a non-vanishing
cosmological constant the conformal infinity is spacelike
(for ${\Lambda>0}$) or timelike (for ${\Lambda<0}$), and the radiative component
of each field depends substantially on the null direction along which $P$ is approached.

The directional dependence of asymptotic fields near such de~Sitter-like or anti--de~Sitter-like
$\scri$ is explicitly found and described. We demonstrate that the corresponding
directional structure of radiation has a universal character that is
determined by the  algebraic (Petrov) type of the field.
In particular, when ${\Lambda>0}$ the radiation  vanishes only along directions
which are opposite to principal null directions.
For ${\Lambda<0}$ the directional dependence is more complicated because it is
necessary to distinguish outgoing and ingoing radiation. Near  such
anti--de~Sitter-like conformal infinity the corresponding directional structures
differ, depending not only on the number and degeneracy
of the principal null directions at~$P$ but also on their specific
orientation with respect to $\scri$.

The directional structure of radiation near (anti--)de~Sitter-like infinities
supplements the standard peeling-off property of spin-$s\,$ fields. This
characterization offers a better understanding of the asymptotic behaviour
of the fields near conformal infinity under the presence of a cosmological constant.

\end{abstract}

\submitto{\CQG}

\newpage

\section*{Table of contents}

\smallskip
\contentsline {section}{\numberline {\ref{sc:introduction}}Introduction}{\pageref{sc:introduction}}
\contentsline {subsection}{\numberline {\ref{ssc:introrad}}On studies of asymptotic behaviour of radiative fields}{\pageref{ssc:introrad}}
\contentsline {subsection}{\numberline {\ref{ssc:outline}}Outline of the present work}{\pageref{ssc:outline}}
\contentsline {section}{\numberline {\ref{sc:infinity}}Conformal infinity and null geodesics}{\pageref{sc:infinity}}
\contentsline {subsection}{\numberline {\ref{ssc:ConfGeom}}Conformal geometry}{\pageref{ssc:ConfGeom}}
\contentsline {subsection}{\numberline {\ref{ssc:ConfInf}}Conformal infinity ${\scri}$ and its character}{\pageref{ssc:ConfInf}}
\contentsline {subsection}{\numberline {\ref{ssc:geodesics}}Null geodesics}{\pageref{ssc:geodesics}}
\contentsline {section}{\numberline {\ref{sc:tetrads}}Various null tetrads}{\pageref{sc:tetrads}}
\contentsline {subsection}{\numberline {\ref{ssc:tetradtrans}}Tetrads and their transformations}{\pageref{ssc:tetradtrans}}
\contentsline {subsection}{\numberline {\ref{ssc:AdjTetr}}The tetrad adjusted to ${\scri}$}{\pageref{ssc:AdjTetr}}
\contentsline {subsection}{\numberline {\ref{ssc:IntTetr}}The interpretation tetrad}{\pageref{ssc:IntTetr}}
\contentsline {subsection}{\numberline {\ref{ssc:AsympIntTetr}}Asymptotic behaviour of the interpretation tetrad}{\pageref{ssc:AsympIntTetr}}
\contentsline {subsection}{\numberline {\ref{ssc:RefTetr}}The reference tetrad and parameterization of null directions}{\pageref{ssc:RefTetr}}
\contentsline {section}{\numberline {\ref{sc:fields}}The fields and their asymptotic structure}{\pageref{sc:fields}}
\contentsline {subsection}{\numberline {\ref{ssc:fieldtrans}}The field components and their transformation properties}{\pageref{ssc:fieldtrans}}
\contentsline {subsection}{\numberline {\ref{ssc:PND}}Principal null directions and algebraic classification}{\pageref{ssc:PND}}
\contentsline {subsection}{\numberline {\ref{ssc:fieldinint}}Field components in the interpreation tetrad}{\pageref{ssc:fieldinint}}
\contentsline {subsection}{\numberline {\ref{ssc:fieldinref}}Asymptotic behaviour of the field components in the reference tetrad}{\pageref{ssc:fieldinref}}
\contentsline {subsection}{\numberline {\ref{ssc:dirstrrad}}Asymptotic directional structure of radiation}{\pageref{ssc:dirstrrad}}
\contentsline {section}{\numberline {\ref{sc:pattern}}Discussion of the directional structure of radiation on ${\scri}$}{\pageref{sc:pattern}}
\contentsline {subsection}{\numberline {\ref{ssc:nullscri}}Radiation on null ${\scri}$}{\pageref{ssc:nullscri}}
\contentsline {subsection}{\numberline {\ref{ssc:peeling}}On the meaning of the peeling-off behaviour}{\pageref{ssc:peeling}}
\contentsline {subsection}{\numberline {\ref{ssc:SpherAngles}}Parametrization of directions by (pseudo-)spherical angles}{\pageref{ssc:SpherAngles}}
\contentsline {subsection}{\numberline {\ref{ssc:spacelikesri}}Radiation on spacelike ${\scri}$}{\pageref{ssc:spacelikesri}}
\contentsline {subsection}{\numberline {\ref{ssc:timelikescri}}Radiation on timelike ${\scri}$}{\pageref{ssc:timelikescri}}
\contentsline {section}{\numberline {\ref{sc:conclude}}Conclusions}{\pageref{sc:conclude}}
\contentsline {section}{\numberline {\ref{apx:expansions}}Asymptotic polyhomogenous expansions}{\pageref{apx:expansions}}
\contentsline {section}{\numberline {\ref{apx:spinors}}Tetrads and fields in spinor formalism}{\pageref{apx:spinors}}
\contentsline {section}{\numberline {}References}{\pageref{sc:references}}

\maketitle

\section{Introduction}
\label{sc:introduction}
Many studies have been devoted to theoretical
investigations  of gravitational waves. The first --- by Einstein
himself --- appeared immediately after the formulation of general
relativity \cite{Einstein:1916b,Einstein:1918}, and was
soon followed by other papers \cite{Weyl:book,Eddington:1923}.
Since then numerous works on gravitational radiation have concentrated on specific
\emph{approximate} (analytic or numerical) analyses of various spatially isolated gravitating sources,
most recently  binary systems, collision and merge of black holes or neutron stars,
supernova explosions, and other possible astrophysical sources.

In \emph{rigorous} treatments within the full Einstein theory, several interesting
classes of exact radiative solutions were found  and investigated in the
late 1950s and the early 1960s --- for example\
\cite{BondiPiraniRobinson:1959,JordanEhlersKundt:1960,EhlersKundt:1962,RobinsonTrautman:1960,RobinsonTrautman:1962,%
BonnorSwaminarayan:1964,Bicak:1968}. For reviews of these contributions to the theory of gravitational radiation
see, e.g., \cite{Stephanietal:book,Bicak:Bonnor,Bicak:Robinson,BicakSchmidt:1989,BonnorGriffithsMacCallum:1994,%
Bicak:1997,Bicak:Ehlers,BicakKrtous:2003}.
Although most of such spacetimes seem to be physically not very realistic, they serve
as useful explicit models and test beds for numerical relativity and other approximations.
Almost simultaneously, general frameworks which allow one to study asymptotic properties
of radiative fields were also developed and applied in now classical works
\cite{Bondi:1960,BondiBurgMetzner:1962,Sachs:1961,Sachs:1962,Burg:1969,NewmanPenrose:1962,%
NewmanUnti:1962,Penrose:1963,Penrose:1964,Penrose:1965}
and elsewhere (see e.g.
\cite{Pirani:1965,Geroch:1977,NewmanTod:1980,PenroseRindler:book,Friedrich:1992,Stewart:book}
for reviews and many references).

Despite this long-standing effort, however, there still remain open
fundamental problems concerning the very concept of gravitational  radiation
in the context of the \emph{full} nonlinear Einstein theory.
No rigorous statements are available which would relate the properties of sufficiently
\emph{general} strong sources to the radiation fields produced.
Also,  a presence of a  \emph{non-vanishing cosmological constant} $\Lambda$
is not compatible with asymptotic flatness that is
naturally assumed in many of the existing analyses. Although important results on
the existence of vacuum solutions with ${\Lambda\not=0}$ have already been obtained
\cite{Friedrich:1998a}, in order to fully understand the properties of gravitational
and electromagnetic radiation in such \vague{de~Sitter-type} or \vague{anti--de~Sitter-type}
spacetimes, further studies are necessary.

As a particular contribution to this task we will here describe
the asymptotic directional behaviour of general fields in
spacetimes which admit any value of the cosmological constant.

\subsection{On studies of asymptotic behaviour of radiative fields in general relativity}
\label{ssc:introrad}
First, we briefly summarize the main  methods  which have been developed
to  characterize rigorously the asymptotic properties of fields in general relativity.
It is not our intention to present an exhaustive and thorough review of
previous works. We only wish to set up a context in which we could
place our present analysis and results.

One fundamental technique for investigating radiative properties of
gravitational and electromagnetic fields at \vague{large distance} from a spatially bounded
source is based on introducing  a suitable \emph{Bondi-Sachs coordinate system} adapted
to \emph{null hypersurfaces}, and expanding the metric functions in inverse
powers of the luminosity distance $r$ which plays the role of an appropriate
\vague{radial} coordinate  parameterizing outgoing null geodesics
\cite{Bondi:1960,BondiBurgMetzner:1962,Sachs:1962,Sachs:1962b}.
In the case of asymptotically flat spacetimes this
framework allows one to introduce the Bondi mass (the total mass of a system as measured
at future null infinity~$\scri$) and momentum, and characterize the time evolution including
radiation in terms of the news functions which are the analogue of the radiative part of
the Poynting vector in electrodynamics. Using these concepts, it is possible
to formulate a balance between the amount of energy radiated by gravitational waves
and the decrease of the Bondi mass of an isolated system.
These pioneering contributions were subsequently refined and generalized
\cite{Burg:1966,Burg:1969,BicakPravdova:1998}, and also extended after
the development of the complex null tetrad formalism and the associated spin coefficient
formalism \cite{NewmanPenrose:1962,NewmanUnti:1962}  which lead to great simplifications
in the expressions, see e.g. \cite{Pirani:1965,NewmanTod:1980,Bicak:Bonnor,Bicak:1997,Bicak:Ehlers}
for reviews. Nevertheless, in these works the analysis of radiative fields assumed that spacetime
is asymptotically flat. This ruled out, for instance, a non-vanishing cosmological
constant $\Lambda$. These methods generally are based on privileged coordinate systems which
are not automatically possible to generalize to the cases when ${\Lambda\not=0}$.

Alternatively, information about the character of radiation can be extracted from the
tetrad components of fields measured along a family of null geodesics approaching~$\scri$.
One can consider only a bundle of such geodesics, as $\scri$ need not exist globally.
The rate of approach to zero of the Weyl or  Maxwell tensor is  given  by the celebrated
\emph{peeling-off theorem} \cite{Sachs:1961,Sachs:1962,NewmanPenrose:1962,GoldbergKerr:1964,Penrose:1965,PenroseRindler:book}.
The component of a spin-$s$  zero-rest-mass field (with respect to a parallelly transported
and suitably normalized interpretation tetrad) proportional to $\afp^{-(j+1)}$,
where $\afp$ is an affine parameter along the null geodesics, ${j = 0,1,\dots,2s}$,
appears to have in general ${2s-j}$ coincident principal null directions.
Consequently, the part of the field that falls off as ${\afp^{-1}}$ exhibits ${2s}$-degeneracy of
principal null directions. It is thus considered as the \emph{radiation field}
because its asymptotic algebraic \vague{null} structure
\cite{Petrov:1954,Pirani:1957,Debever:1959,Penrose:1960,Stephanietal:book,PenroseRindler:book}
locally resembles that of standard plane waves \cite{BondiPiraniRobinson:1959}. The gravitational
or electromagnetic field thus represents outgoing radiation if the dominant component
of the Weyl or Maxwell tensor, conveniently expressed in the Newman-Penrose formalism
\cite{NewmanPenrose:1962,NewmanUnti:1962,JanishNewman:1965} as  quantities $\WTP{}{4}$ or $\EMP{}{2}$, respectively,
is non-vanishing. This component manifests itself through typical transverse effects on nearby test particles
\cite{Sachs:1962,Szekeres:1965,BicakPodolsky:1999b}. Such a characterization of the radiative field
remains valid also in more general spacetimes  because  the peeling-off property
holds even for a non-vanishing ${\Lambda}$ (the precise meaning of the peeling-off behaviour
of fields will be discussed below in the main text, see sections~\ref{ssc:peeling} and \ref{sc:conclude}).

Another major step made by Penrose \cite{Penrose:1963,Penrose:1964,Penrose:1965,Penrose:1968}
(see \cite{PenroseRindler:book} for a comprehensive overview) was his {\em coordinate-independent}
(geometric) approach to the definition of radiation for massless fields based on  the
\emph{conformal treatment of infinity}.  The Penrose  technique enables one to apply
methods of local differential geometry near conformal infinity $\scri$ (also referred to as \vague{scri})
which is defined as the boundary ${\om=0}$ of the physical spacetime manifold ${(\mfld,\mtrc)}$
in the conformally related \vague{unphysical}  spacetime manifold $({\cmfld,\cmtrc)}$,
${\cmtrc=\om^2\mtrc}$ (see section~\ref{sc:infinity}). Properties of radiation fields in $\mfld$
can thus be studied by analyzing conformally (i.e., isotropically) rescaled fields on $\scri$ in
the compactified manifold $\cmfld$.  For  asymptotically flat spacetimes, $\scri$ is
a  smooth null hypersurface in $\cmfld$ generated by the endpoints of null geodesics. In this case
it is possible to define in a geometric way  the Bondi mass, to derive the peeling-off
property, or to characterize the Bondi-Metzner-Sachs  group of asymptotic symmetries
\cite{BondiBurgMetzner:1962,Sachs:1962b,Penrose:1963,WinicourTamburino:1965,TamburinoWinicour:1966}.
In particular, one can evaluate gravitational radiation propagating along a given null geodesic which
is described by the $\WTP{}{4}$ component of the Weyl tensor projected on a parallelly
transported complex null tetrad.  The crucial point is that such a tetrad is (essentially)
\emph{determined   uniquely} by the conformal geometry, see  \cite{PenroseRindler:book}.
Moreover, the Penrose covariant approach can be naturally
applied also to spacetimes which include the cosmological
constant \cite{Penrose:1964,Penrose:1965,PenroseRindler:book,Friedrich:2002}. This is quite remarkable, since there
is no analogue of the news function in the presence of $\Lambda$ \cite{AshtekarMagnon:1984,AshtekarDas:2000}
(for a  comparison of the Bondi-Sachs and Penrose approaches see, e.g.,
\cite{TamburinoWinicour:1966,Winicour:1968,Persides:1979,Schmidt:1987,ChruscielJezierskiMacCallum:1998,%
TafelPukas:2000,Tafel:2000}).

The above mentioned  analysis lead Penrose to the elegant idea of \emph{(weakly) asymptotically
simple spacetimes} --- those having a \emph{smooth} $\cmtrc$ and $\om$ on $\scri$  (see, e.g.,
\cite{Penrose:1963,Penrose:1965,HawkingEllis:book,PenroseRindler:book,Wald:book1984,Friedrich:1992,Stewart:book,Frauendiener:2004}
and references  therein for the precise definition). Because it entails a certain fall-off behaviour of
the physical metric near $\scri$, this is a fruitful rigorous concept for studying asymptotic radiation
properties of isolated systems in general  relativity.  It follows from the important recent works \cite{Friedrich:1981,Friedrich:1986a,Friedrich:1983,Friedrich:1988,Friedrich:1991,Friedrich:1995,Friedrich:1998c},
and also \cite{CutlerWald:1989,ChristodoulouKlainerman:book,KlainermanNicolo:1999,KlainermanNicolo:2003,%
AnderssonChruscielFriedrich:1992,AnderssonChrusciel:1996,%
Kannar:1996,Corvino:2000,CorvinoSchoen:2003,ChruscielDelay:2002,Friedrich:2003},
that there indeed \emph{exist} large classes of exact --- though not given in explicit forms ---
solutions to Einstein's field equations which globally satisfy the required regularity conditions
on $\scri$ (see \cite{Friedrich:1998a,Friedrich:1998b,Rendall:2002,BicakKrtous:2003,Frauendiener:2004} for a review).
Let us emphasize that, until now, the only \emph{explicitly} known exact metrics satisfying
the Penrose's asymptotic conditions (although not globally since there are at least four \vague{points}
at $\scri$ which are singular) are boost-rotation symmetric
spacetimes representing uniformly accelerated \vague{sources} or black holes
\cite{Bicak:1968,Bicak:Bonnor,Bicak:Robinson,BicakSchmidt:1989,Bicak:1997,BicakPravdova:1998,%
Bicak:Ehlers,Pravdovi:2000,AshtekarDray:1981}. Nevertheless, the original Penrose conjecture of
asymptotic simplicity may appear to be too restrictive in general. More recent studies
have indicated that \emph{generic} Cauchy data fail to be smoothly extendable to the conformal boundary \cite{AnderssonChrusciel:1993,AnderssonChruscielFriedrich:1992,AnderssonChrusciel:1994,AnderssonChrusciel:1996}, see also \cite{ChristodoulouKlainerman:book,Winicour:1985,Damour:1986,Friedrich:2003,Valiente-Kroon:2004}.
More general spacetimes with \emph{polyhomogeneous} $\scri$ were thus studied in
\cite{Chruscieletal:1995,ChruscielJezierskiMacCallum:1998,Valiente-Kroon:2002} and in other works, for which the metric
$\cmtrc$ admits an asymptotic expansion in terms of ${\,r^{-j}\log^i r\,}$ rather than ${r^{-j}}$.
This new setup naturally extends the Bondi-Sachs-Penrose approach. For example, when ${\Lambda=0}$
the Bondi mass still remains well defined at polyhomogeneous $\scri$, and it is a
non-increasing function of retarded time as in \cite{Trautman:1958,BondiBurgMetzner:1962,Sachs:1962}.
For the class of polyhomogeneous vacuum metrics the asymptotic symmetry group is the standard BMS group,
and the peeling-off property of the curvature tensor is the same as the one for smooth metrics
\cite{Sachs:1962} \emph{up to the terms of order} ${r^{-(2+\epsilon)}}$, ${0\le\epsilon<1}$, but
the term ${\sim r^{-3}\log r}$ also appears.  Further studies of polyhomogeneity for zero-rest-mass fields,
such as the existence of conserved quantities (NP constants) at $\scri$
\cite{NewmanPenrose:1965,NewmanPenrose:1968,ExtonNewmanPenrose:1969,Robinson:1969,%
GlassGoldberg:1970,PenroseRindler:book} can be found, e.g., in \cite{Valiente-Kroon:1998,Valiente-Kroon:2000b}.

Let us now concentrate on the main \emph{differences} between the asymptotically flat spacetimes
and those with a non-vanishing cosmological constant $\Lambda$.
Interestingly, specific new features appear in the case of asymptotically
\vague{de~Sitter-like} (${\Lambda>0}$) or \vague{anti--de~Sitter-like}
(${\Lambda<0}$) solutions for which the conformal infinity~$\scri$ is, respectively,
spacelike or timelike. As Penrose observed and repeatedly emphasized in his early
works  \cite{Penrose:1964,Penrose:1965,Penrose:1967}, the concept
of radiation for  massless fields turns out to be \vague{less invariant}
in cases when $\scri$ does not have a null character, see section 9.7 of \cite{PenroseRindler:book}. Namely, it emerges as necessarily
\emph{direction dependent} since the choice of the appropriate
null tetrad, and thus the radiative component $\WTP{}{4}$ of the
field, may differ for different null geodesics reaching the same
point on $\scri$. This is, for example, demonstrated by the fact that with a non-vanishing $\Lambda$
even fields of \vague{static}, non-accelerated sources have a non-vanishing radiative component along a
\emph{generic} (\vague{non-radial}) direction, as it was shown for test charges in
\cite{BicakKrtous:2002} or for Reissner-Nordstr\"om black holes in
\cite{KrtousPodolsky:2003,PodolskyOrtaggioKrtous:2003} in a (anti--)de~Sitter universe.

For ${\Lambda\not=0}$, the non-null character of conformal infinity $\scri$  also plays a
fundamental role in the formulation of the initial value problem. As mentioned above,
quite surprisingly  the
\emph{global} existence has been established of asymptotically  simple vacuum solutions
(with a smooth $\scri$) which differ on an arbitrary given Cauchy surface by a
finite but sufficiently small amount from de~Sitter data \cite{Friedrich:1986a,Friedrich:1998a},
while an analogous result for data close to Minkowski (${\Lambda = 0}$)
is still under investigation (see \cite{Friedrich:1998a,Friedrich:1998b,BicakKrtous:2003}
for more details). Thus, many vacuum
asymptotically simple spacetimes  with the de~Sitter-like $\scri$ do exist.
However, a spacelike $\scri$ as occurs in this case implies the existence of cosmological and particle horizons
for geodesic observers, which results in insufficiency of purely retarded massless
fields  --- advanced effects must necessarily be present. For example, the electromagnetic
field produced by sources cannot be prescribed freely because the Gauss constraint
has to be satisfied at $\scri^-$ (or $\scri^+$).  This phenomenon has been
demonstrated explicitly \cite{BicakKrtous:2001} by analyzing test electromagnetic fields
of uniformly accelerated charges on de~Sitter background. On the other hand, it is
well-known that for a timelike $\scri$, which occurs when ${\Lambda<0}$,
the spacetimes are not globally hyperbolic, and
one is necessarily led to a kind of \vague{mixed initial boundary value
problem}, see, e.g., \cite{Avisetal:1978,Hawking:1983,HenneauxTeitelboim:1985,Friedrich:1995}.
The data need to be given on a spacetime slice extending to $\scri$ and also on $\scri$ itself.
The rigorous concept of gravitational and electromagnetic radiation is thus
much less clear in situations when ${\Lambda\not=0}$.

Here we will analyze mainly the \emph{directional structure} of radiative fields.
This structure is significantly different for null and spacelike/timelike conformal infinities.
In the case of asymptotically flat spacetimes, the dominant radiative component of the field
at any point $P$ at null infinity $\scri$ is essentially \emph{unique}.
One can however approach a point~$P$ from \emph{infinitely many different} null directions,
and if  $\scri$ has a spacelike or timelike character
it is not a~priori clear how the radiation components of the fields
in the corresponding interpretation tetrads depend on a specific direction.

Such a directional dependence was explicitly found and described for the first time in
the context of the \emph{test electromagnetic field} generated by a pair of uniformly  accelerated
point-like charges in the de~Sitter background \cite{BicakKrtous:2002,BicakKrtous:BIS}.
In particular, it was demonstrated that there always exist two special directions ---
those opposite  to the direction from the sources --- along which the radiation vanishes.
For all other directions the radiation field is non-vanishing. This is described by an explicit
formula which completely characterizes its angular dependence (see \cite{BicakKrtous:2002}).

Subsequently, we have carefully analyzed the \emph{exact} solution of the
Einstein-Maxwell equations which generalizes the classic $C$-metric
(see e.g. \cite{EhlersKundt:1962,KinnersleyWalker:1970,BicakPravda:1999,Pravdovi:2000}
for reviews and references) to admit a cosmological constant
\cite{PlebanskiDemianski:1976,PodolskyGriffiths:2001,Podolsky:2002,DiasLemos:2003a,DiasLemos:2003b}.
For ${\Lambda>0}$ it represents a pair of uniformly accelerated possibly charged
black holes in  de~Sitter-like universe. In \cite{KrtousPodolsky:2003}
we demonstrated that the corresponding electromagnetic field  exhibits exactly
the \emph{same} asymptotic radiative behaviour at the spacelike conformal
infinity $\scri$ as for the test fields \cite{BicakKrtous:2002} of accelerated
charges. Moreover, we found and explicitly described the specific analogous
directional structure of the \emph{gravitational radiation field}, and we proved that
the directional pattern of radiation is adapted to the principal null directions
of this Petrov type~D spacetime.

Elsewhere \cite{PodolskyOrtaggioKrtous:2003} we investigated the
asymptotic behaviour of fields corresponding to the $C$-metric with ${\Lambda<0}$,
i.e.  the directional dependence of radiation  generated by accelerated
black holes in an anti--de~Sitter universe. Some fundamental differences
from the case ${\Lambda>0}$ occur since the conformal infinity $\scri$
now has a timelike character. In fact, the whole structure of the spacetime
is more complex and new phenomena also arise: $\scri$ is divided by Killing
horizons into several domains with a different structure of
principal null directions --- in these domains the directional structure of radiation
is thus different. The radiative field vanishes along directions which are mirror images
of the principal null directions with respect to $\scri$.
Moreover, ingoing and outgoing radiation has to be treated separately.

These studies of particular exact radiative models with a non-vanishing ${\Lambda}$
gave us a sufficient insight  necessary to understand the
asymptotic behaviour of  \emph{general} fields near spacelike or
timelike conformal infinities. The directional dependence of gravitational
and electromagnetic radiation is given mainly by \emph{spacetime geometry}, namely
by character of ${\scri}$ and
by specific orientation and degeneracy of principal null directions
at infinity.

In \cite{KrtousPodolskyBicak:2003} we demonstrated that  the directional structure
of radiation close to a de~Sitter-like infinity  has a \emph{universal character}
that is determined by the \emph{algebraic type} of the fields. For example, the radiation
completely vanishes along spatial directions on $\scri$ which are antipodal to
principal null directions. In the following  work \cite{KrtousPodolsky:2004a}
we investigated the complementary  situation when $\Lambda<0$.
Although the idea is similar to the previous case, the asymptotic  behaviour
of fields turns out to be more complicated because  $\scri$ is timelike, and thus
admits a \vague{richer structure} of possible radiative patterns.

\subsection{Outline of the present work}
\label{ssc:outline}
It is the purpose of this review to present these results --- concerning the  asymptotic directional
structure of general fields near conformal infinity $\scri$ of any type --- in a synoptic, compact
and unified form. The paper is organized as follows. First, in section~\ref{sc:infinity} we
summarize basic concepts of conformal geometry and their relation to quantities in a physical
spacetime. In particular, we introduce the conformal infinity~$\scri$, correlate its character with the sign
of the cosmological constant $\Lambda$, and we investigate the correspondence between null
geodesics in physical and conformal spacetimes. Section~\ref{sc:tetrads} is devoted to careful analysis
of various orthonormal and null tetrads which are key ingredients in our subsequent study of the
asymptotic behaviour of fields. We define an interpretation tetrad which is parallelly propagated
along a null geodesic, and we demonstrate that --- after performing a specific boost --- it becomes
asymptotically adjusted to $\scri$ (i.e., naturally normalized and adapted to normal and tangent
directions at a given point of conformal infinity). This fact becomes crucial in section~\ref{sc:fields} in
which we explicitly evaluate the components of general gravitational, electromagnetic, or any
spin-$s$ field in the interpretation tetrad near conformal infinity. For zero-rest-mass
fields the dominant component decays as ${\afp^{-1}}$ (where $\afp$ is the affine
parameter of a null geodesic) and thus represents radiation.

The complete
expression \eqref{FieldInterpComplete} fully chracterizes the asymptotic behaviour of the field near any $\scri$,
including the directional structure, i.e. the dependence on the direction of the
geodesic along which a point ${P\in\scri}$ is approached as ${\afp\to\infty}$. The final
section~\ref{sc:pattern} contains a detailed discussion of the result. In asymptotically Minkowskian spacetimes
with ${\Lambda=0}$ for which $\scri$ has a null character the directional dependence completely vanishes.
The presence or absence of the ${\afp^{-1}}$ component of the field can thus be used as an
invariant characterization of radiation. In this context we also elucidate the precise meaning of the
peeling-off behaviour. For ${\Lambda\not=0}$ the asymptotic structure of fields is more complicated because
the dominant component depends substantially on the direction of a null geodesic along which
$P$ is approached. We introduce a convenient parametrization of such directions near a spacelike or
timelike $\scri$ which occur in spacetimes with ${\Lambda>0}$ or ${\Lambda<0}$, respectively.
Finally, we describe in detail the asymptotic directional structure  of radiative fields
near such de~Sitter-like or anti--de~Sitter-like conformal infinities. It is proved to be
essentially determined by the algebraic type of the field, namely by the number, degeneracy, and
specific orientation of the principal null directions at point ${P\in\scri}$.

Two appendices are included. In appendix~\ref{apx:expansions}
we present expansions of the conformal factor $\om$ and the conformal affine parameter $\cafp$
in terms of the physical affine parameter $\afp$ of a null geodesic. We demonstrate
their polyhomogeneous character. We also investigate the conditions under which the two main
vectors of the interpretation tetrad become coplanar with the normal to $\scri$.
Appendix~\ref{apx:spinors} summarizes the description of  spin-$s$ fields,
tetrads and their Lorentz transformations in spinor formalism.

\section{Conformal infinity and null geodesics}
\label{sc:infinity}
In this section we recall some basic concepts and properties concerning the geometry
of a physical spacetime and its conformally related counterpart which will be
necessary for our subsequent analysis. Many of these concepts can be found in
standard literature, e.g., in \cite{Stephanietal:book,PenroseRindler:book,Wald:book1984}.
However, we summarize them for convenience and to introduce our notation.

\subsection{Conformal geometry}
\label{ssc:ConfGeom}
We wish to study spacetimes which locally admit conformal infinity.
According to general formalism \cite{Penrose:1963,Penrose:1965,Geroch:1977,PenroseRindler:book,Wald:book1984},
such $n$-dimensional manifold $\mfld$ with physical metric~$\mtrc$ can be embedded into
a larger \defterm{conformal manifold} $\cmfld$ with \defterm{conformal metric}~$\cmtrc$
via a conformal transformation
\begin{equation}\label{ConfMetric}
  \cmtrc=\om^2\mtrc  \period
\end{equation}
Obviously, the spacetimes $(\mfld,\mtrc)$ and
$(\cmfld,\cmtrc)$ have identical local causal structure (the same light cones).
The conformal factor ${\om}$ is assumed to be positive in $\mfld$, and vanishes on the boundary
of $\mfld$ in $\cmfld$. Such a boundary ${\om=0}$ is called \defterm{conformal
infinity}~$\scri$. Let us note that in the following it is not necessary
to require a global existence of~$\scri$. However, we assume that $\om$
is smooth near $\scri$, and we impose suitable regularity conditions for the conformal
metric. Specifically, we assume that the conformal factor is sufficiently smooth
along null geodesics approaching $\scri$ in $(\cmfld,\cmtrc)$,
cf.\ equation~\eqref{OmegaAnalyt} in this section.

To indicate explicitly which of the above metrics is used
for raising  indices, we introduce $\mtrc^{ab}$ as the inverse of
$\mtrc_{ab}$, and  $\cmtrc^{ab}$ as the inverse of $\cmtrc_{ab}$.

The derivative operator of $\cmtrc$ is related to that of $\mtrc$.
The relation between the derivative $\ccovd$ associated with $\cmtrc$,
and $\covd$ associated with $\mtrc$ is (see, e.g., \cite{Wald:book1984})
\begin{equation}\label{DerivRelation}
  \ccovd_a\tens{v}^c=\covd_a\tens{v}^c+\covddif_{ab}^c\tens{v}^b \comma\quad
  \covddif_{ab}^c=\om^{-1}(\delta_a^c\grad_b\om+\delta_b^c \grad_a \om
    -\mtrc_{ab}\mtrc^{cd}\grad_d \om)  \period
\end{equation}
This implies relations between the curvature associated
with $\ccovd$, and the curvature associated with $\covd$. By
contracting the formula for the Riemann tensor
we obtain relations
between the conformal and physical Ricci tensors,
\begin{eqnarray}\label{RicciRelation}
 \cRic_{ab}&=&\Ric_{ab}-(n-2)\om^{-1}\covd_a\grad_b\om
   -\om^{-1}\mtrc_{ab}\dalamb\om \nonumber\\
   && \qquad+2(n-2)\om^{-2}\grad_a \om\,\grad_b  \om
     -(n-3)\om^{-2}\mtrc_{ab}\mtrc^{cd}\grad_c \om \,\grad_d \om    \commae
\end{eqnarray}
where ${\dalamb=\mtrc^{ab}\covd_a\covd_b}$, and the scalar curvatures,
\begin{equation}\label{scalarcurvRelation}
 \cscR=\om^{-2}\scR-2(n-1)\om^{-3}\dalamb\om
   -(n-1)(n-4)\om^{-4}\mtrc^{ab} \grad_a \om \,\grad_b \om     \period
\end{equation}
The Weyl tensor  is unchanged by a conformal
transformation,
\begin{equation}\label{WeylRelation}
 \cWT_{abc}^{\quad\,d}=\WT_{abc}^{\quad\,d}   \period
\end{equation}

\subsection{Conformal infinity ${\scri}$ and its character}
\label{ssc:ConfInf}
\begin{figure}
\begin{center}
\includegraphics[scale=0.75]{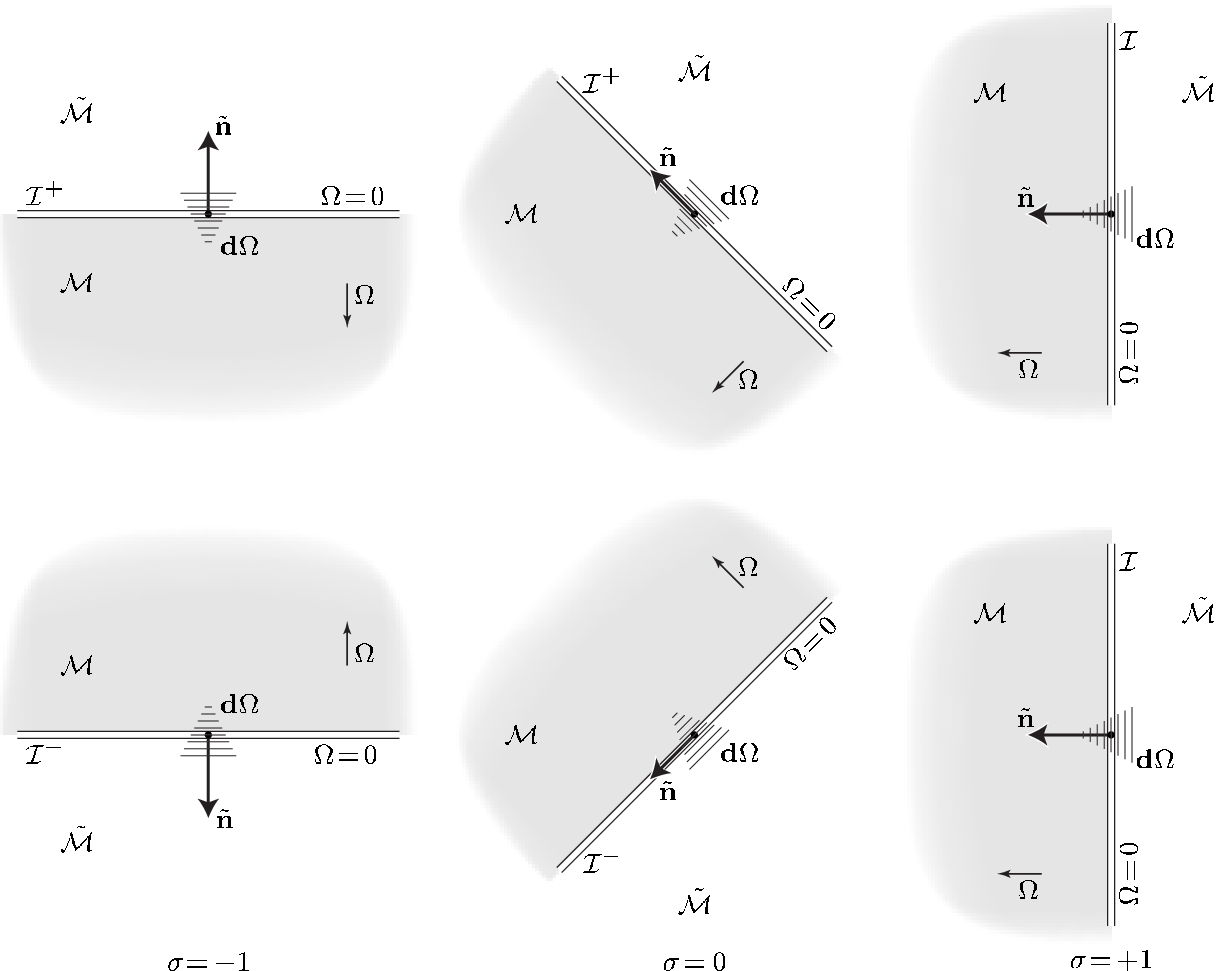}
\end{center}
\caption{\label{fig:scri}%
The local character of the conformal infinity $\scri$ (situated on
the boundary ${\om=0}$ of $\mfld$ in $\cmfld$) is determined by the norm
$\nsgn$ of vector $\cnorm$ normal to $\scri$. For ${\nsgn=-1, 0}$, or ${+1}$
the infinity $\scri$ is spacelike, null, or timelike, respectively. When
${\nsgn=-1}$ or ${\nsgn=0}$, the  future conformal infinity $\scri^+$ and the
past conformal infinity $\scri^-$ can be distinguished; the corresponding
diagrams are drawn in the upper and lower part of the figure.
For ${\nsgn=+1}$ the future and past infinities of null geodesics are
the same, and therefore the diagrams are identical.}
\end{figure}

The conformal infinity is localized by the condition  ${\om=0}$. Its
character is determined by the gradient $\grad\om$ on $\scri$ --- it can be timelike,
null, or spacelike. We introduce a normalized vector $\cnorm$ which is
\emph{normal to the conformal infinity} $\scri$,
\begin{equation}\label{NormVect}
  \cnorm^a=\clapse\,\cmtrc^{ab}\,\grad_b\om\comma\quad
  \cmtrc_{ab}\cnorm^a\cnorm^b=\nsgn\comma\quad
  \nsgn=-1,0,+1\period
\end{equation}
For ${\nsgn=\pm1}$ the conformal \vague{lapse} function ${\clapse>0}$ is given by
$\clapse=|\cmtrc^{ab}\grad_a \om \,\grad_b \om|^{-1/2}$,
and ${\cmtrc = \nsgn\clapse^2\,\grad\om{}^2+\scricmtrc}$, where
$\scricmtrc$ is a restriction of $\cmtrc$ on $\scri$.
For ${\nsgn=0}$ the \vague{lapse} ${\clapse}$ is chosen arbitrarily.
The normalization factor $\nsgn$
determines the character of the conformal infinity, namely
\begin{equation}
\nsgn=
\begin{cases}
   -1: &       \quad \text{$\scri\,$ is \emph{spacelike},}  \cr
   \hfill 0: & \quad \text{$\scri\,$ is \emph{null},} \cr
   +1: &       \quad \text{$\scri\,$ is \emph{timelike}.} \cr
\end{cases}
\label{DefinitionSigma}\end{equation}
This is illustrated in figure~\ref{fig:scri}.

In fact, we can explicitly evaluate the \vague{lapse} $\clapse$.
Transforming ${\dalamb}$ to  ${\cdalamb=\cmtrc^{ab}\ccovd_a\ccovd_b}$ in \eqref{scalarcurvRelation} using
relation \eqref{DerivRelation}, we obtain
\begin{equation}\label{ConfscalarcurvRelation}
\cmtrc^{ab} \grad_a \om \,\grad_b \om = -\frac{\scR}{n(n-1)}
  +\om\, \biggl(\frac{2}{n}\,\cdalamb\om+\frac{\om\cscR}{n(n-1)}\biggr) \period
\end{equation}
By contracting the Einstein field equations
\begin{equation}\label{EinstEq}
\Ric-{\textstyle\frac{1}{2}}\scR\,\mtrc+\Lambda\,\mtrc=\kap\,\TEM  \commae
\end{equation}
we get ${\scR=\frac{2}{n-2}(n\Lambda-\kap\, T)}$. Assuming a vanishing
trace $T$ of the energy-momentum tensor, which is valid in vacuum,
pure radiation, or electrovacuum (${n=4}$) spacetimes, equation \eqref{ConfscalarcurvRelation}
on $\scri$  implies
\begin{equation}\label{conflapse}
  \nsgn\,\clapse\big|_\scri^{-2}=\cmtrc^{ab}\grad_a \om \,\grad_b \om \big|_\scri
   = -\frac{2\Lambda}{(n-1)(n-2)} \comma
  \quad\text{i.e.,}\quad \nsgn=-\sign{\Lambda} \period
\end{equation}
The character of the conformal infinity is thus correlated with the sign of the
cosmological constant. For ${\nsgn\ne0}$ we also
obtain that the \emph{\vague{lapse} is constant on $\scri$}, ${\clapse|_\scri=\scale>0}$,
where $\scale$ is a typical length, which for spacetime dimension ${n=4}$ is ${\scale=\sqrt{|3/\Lambda|}}$.
For (anti--)de~Sitter spacetime, the scale $\scale$ represents its characteristic radius.
When   ${\sigma=0}$ (${\Lambda=0}$)  we choose $\clapse$ to be
an arbitrary constant on $\scri$. The normalized timelike/null/spacelike
vector $\cnorm$ given by \eqref{NormVect}
which is normal to conformal  infinity $\scri$ is thus set uniquely.

Analogously we introduce a vector $\norm$ normal to ${\om=\text{const.}}$ in
the physical spacetime ${(\mfld,\mtrc)}$ such that
\begin{equation}\label{NormNorm}
\mtrc_{ab}\norm^a\norm^b=\nsgn \commae
\end{equation}
which implies the relation
\begin{equation}\label{NormCnormRel}
\norm=\om\,\cnorm \period
\end{equation}
Strictly speaking, it is not possible to introduce the vector ${\norm}$
normalized in the physical geometry ${(\mfld,\mtrc)}$
at ${\scri}$ directly. The conformal infinity \emph{does not} belong to the physical
spacetime, and even if we extend the manifold ${\mfld}$ into the
conformal manifold ${\cmfld}$, the physical metric ${\mtrc}$ is not
well defined on ${\scri}$: it is related to the conformal metric ${\cmtrc}$
by the factor ${\om^{-2}}$, see \eqref{ConfMetric}, which becomes infinite.

We could try to extend the definition of the physical metric (and other tensors
related to physical spacetime) up to the infinity ${\scri}$ using some limiting
procedure, e.g., using its expansion along curves approaching ${\scri}$.
However, the \vague{infinite ratio} between ${\mtrc}$ and ${\cmtrc}$
still posses problems. Physically defined vectors
transported in a natural way to ${\scri}$ are rescaled to zero when
measured in conformal geometry, and vectors from the conformal tangent space
at ${\scri}$ have an infinite length if measured using the physical metric.
The tangent space of the conformal manifold at ${\scri}$
is infinitely \vague{blown up} with respect to the tangent space of the physical manifold
defined at the conformal infinity by a suitable limiting procedure.
Nevertheless, one can deal with such infinite scaling using the conformal technique:
its important feature is that the conformal rescaling is \emph{isotropic} ---
it rescales all directions in the same way.
If the rescaling enters expansions of physical tensor quantities only
as a common factor which is some power of ${\om}$,
it is \vague{well controlled}: we can associate with any physical quantity
a conformal quantity rescaled by a proper power of ${\om}$ which is
correctly defined at conformal infinity, independent
of a direction along which ${\scri}$ is approached.

The conformal geometry and the definition of the conformal infinity can
thus be understood as a convenient way to deal with tensors at infinity
--- any physical quantity can be  \vague{translated} to its conformal
counterpart which is well defined at ${\scri}$. However, it can
be convenient sometimes to speak directly about physical quantities
at ${\scri}$ and we will do so if the \vague{translation} to the conformal picture is clear.
Exactly in this sense, we can speak about the physical normal vector ${\norm}$ at ${\scri}$,
even if it is related to the well defined conformal normal ${\cnorm}$ by
relation \eqref{NormCnormRel} which is degenerate on ${\scri}$.
Similarly, in the next section we use a null tetrad
adjusted to the infinity ${\scri}$ normalized in the physical geometry.

Let us note however that one has to be careful with asymptotic expansions
if these are not isotropic, i.e., if they rescale one direction
by an \vague{infinite amount} as compared with other directions.
This will be the case of, for example, interpretation
null tetrad parallelly transported to ${\scri}$ as discussed in
section~\ref{ssc:AsympIntTetr}. Various components of physical tensors with respect
to the interpretation tetrad thus will not rescale in the same way and
their behaviour has to be studied more carefully, cf.\ section~\ref{ssc:fieldinint}.

\subsection{Null geodesics}
\label{ssc:geodesics}
Now we consider geodesics in the physical spacetime $(\mfld,\mtrc)$ and
we  relate them to geodesics in the conformal
spacetime $(\cmfld,\cmtrc)$. It follows from \eqref{DerivRelation}
that \emph{null geodesics are conformally invariant}, i.e.,
null geodesics $\geod(\afp)$ with respect to $\covd$ coincide with null
geodesics $\cgeod(\cafp)$ with respect to $\ccovd$. The affine parameter
$\cafp$ for geodesics in conformal spacetime is related to the affine parameter
$\afp$ for geodesics in physical spacetime by
\begin{equation}\label{RelatAfpCafp}
\frac{d\cafp}{d\afp}=\om^2\comma \quad
\text{i.e.},\quad
\frac{D\geod}{d\,\afp}=\om^2\frac{D\cgeod}{d\,\cafp}
\commae
\end{equation}
(we fix a  trivial factor corresponding to constant rescaling of $\cafp$ to unity).

Without loss of generality we take the affine parameter $\cafp$ such that
${\cafp=0}$ at conformal infinity $\scri$. Therefore, as ${\cafp\to0}$ the
null geodesic $\cgeod(\cafp)$ in  conformal spacetime approaches a specific
point $P\in\scri$, i.e., ${\cgeod(0)=P}$. Such a geodesic can be either
\defterm{outgoing} or \defterm{ingoing} with respect to
 physical spacetime $\mfld$:
\begin{equation}\label{InOutGeod}
\frac{D\cgeod^a}{d\,\cafp}\,\grad_a\om\bigg|_\scri
    \equiv\frac{d\,\om}{d\,\cafp}\bigg|_\scri=-\EPS \commae
\end{equation}
where
\begin{equation}
\EPS=
\begin{cases}
   +1: &       \quad \text{for \emph{\,outgoing geodesics},}\quad \cafp<0 \text{\, in } \mfld, \cr
   -1: &       \quad \text{for \emph{\,ingoing geodesics},\,}\ \quad \cafp>0 \text{\, in } \mfld. \cr
\end{cases}
\label{DefinitionEpsilon}\end{equation}
By this condition, the normalization of the affine parameter ${\cafp}$ is fixed uniquely,
including the orientation of the null geodesic $\cgeod(\cafp)$.
As we have already mentioned, we assume smoothness of the conformal factor along
$\cgeod(\cafp)$, hence we may expand ${\om}$ in powers of $\cafp$ near $\scri$. Taking into
account that ${\cafp|_\scri=0}$, and equation \eqref{InOutGeod}, we have
\begin{equation}\label{OmegaAnalyt}
\om=-\EPS\,\cafp+\om_2\,\cafp^2+\ldots \commae
\end{equation}
with ${\om_2}$ constant.
Substituting into equation \eqref{RelatAfpCafp}, straightforward integration
leads to the relation between the physical and conformal affine parameters
\begin{equation}\label{AfpOfCafp}
\afp=-\frac{1}{\cafp}\, \bigl(1-2\EPS\, \om_2\,\cafp\ln|\cafp|-\afp_0\,\cafp+\ldots\bigr) \period
\end{equation}
Here ${\afp_0}$ is a constant of integration.
Consequently, near $\scri$ we obtain in the leading order that
${\cafp\lteq-\afp^{-1}}$ and ${\om\lteq\EPS\,\afp^{-1}}$.
The null geodesic $\geod(\eta)$ thus reaches the point ${P\in\scri}$
for an \emph{infinite} value of the affine parameter $\afp$,
namely ${\geod(\EPS\infty)=P}$.

The leading term in the expansions \eqref{OmegaAnalyt} and \eqref{AfpOfCafp}
is sufficient for all calculations throughout this paper. However,  for other purposes
it can be useful to express these expansions up to the next
order. It is not simple to invert the expansion \eqref{AfpOfCafp}
due to the presence of the logarithmic term.
In appendix~\ref{apx:expansions} we demonstrate that (cf.\ equation~\eqref{apx:CafpOfAfp})
\begin{equation}\label{CafpOfAfp}
\cafp=-\frac{1}{\afp}\, \bigl(1-2\EPS\om_2\,\afp^{-1}\ln|\afp|+\afp_0\,\afp^{-1}+\ldots\bigr)\commae
\end{equation}
and  (cf.\ equation~\eqref{apx:OmegaOfAfp})
\begin{equation}\label{OmegaInAfp}
\om=\EPS\,\afp^{-1}+\bigl(-2\om_2\,\ln|\afp|+\EPS\afp_0+\om_2\bigr)\,\afp^{-2}+\ldots\period
\end{equation}
Nevertheless, the logarithmic terms in these expansions disappear provided Penrose's
\defterm{asymptotic Einstein condition} (cf.\ equation 9.6.21 of \cite{PenroseRindler:book}),
\begin{equation}\label{AsymEinstCond}
\ccovd_b\grad_a\om \lteq {\textstyle\frac{1}{4}} \cmtrc_{ab}\cdalamb\om\commae
\end{equation}
is satisfied, cf.\ equation \eqref{om2explicite3}.

\section{Various null tetrads}
\label{sc:tetrads}

We now wish to investigate the behaviour of fields near conformal infinity in
standard general relativity in four dimensions (${n=4}$).
For this purpose we  introduce the normalized \vague{interpretation} tetrad
which is parallelly transported along null geodesics $\geod(\eta)$
approaching $\scri$. To achieve this  we will employ several orthonormal
and null tetrads which will be distinguished by specific labels in subscripts.

\subsection{Tetrads and their transformations}
\label{ssc:tetradtrans}
We denote the vectors of an \defterm{orthonormal tetrad}  as
${\tG,\,\qG,\,\rG,\,\sG}$, were $\tG$ is a unit timelike vector, typically chosen
to be future oriented, and the  remaining three are spacelike unit vectors.
With this tetrad we associate a \defterm{null tetrad}
of null vectors ${\kG,\,\lG,\,\mG,\,\bG}$, such that
\begin{equation}\label{NormNullTetr}
\begin{aligned}
  \kG &= \textstyle{\frac1{\sqrt{2}}} (\tG+\qG)\comma&
  \lG &= \textstyle{\frac1{\sqrt{2}}} (\tG-\qG)\commae\\
  \mG &= \textstyle{\frac1{\sqrt{2}}} (\rG-i\,\sG)\comma&
  \bG &= \textstyle{\frac1{\sqrt{2}}} (\rG+i\,\sG)\commae
\end{aligned}
\end{equation}
where the only non-vanishing scalar products are
\begin{equation}\label{NullTetrNorm}
   \mtrc_{ab}\kG^a\lG^b = -1\comma
   \mtrc_{ab}\mG^a\bG^b = 1 \period
\end{equation}
Similarly, we introduce  a \defterm{conformal null tetrad} ${\kC,\,\lC,\,\mC,\,\bC}$
in conformal spacetime $\cmfld$ normalized by the conformal metric
$\cmtrc$ as $\,{\cmtrc_{ab}\kC^a\lC^b = -1}$, $\,{\cmtrc_{ab}\mC^a\bC^b = 1}$,
which is associated with \defterm{conformal orthonormal tetrad}
${\tC,\,\qC,\,\rC,\,\sC}$.

Transformations between orthonormal tetrads (and corresponding null tetrads)
form the Lorentz group. In the context of null tetrads it is convenient to consider four
simple transformations from which any Lorentz transformation
can be generated \cite{Stephanietal:book}: \defterm{null rotation with ${\kG}$ fixed},
parametrized by ${L\in\complexn}$,
\begin{equation}\label{kfixed}
  \kG = \kO\commae\qquad
  \lG = \lO + \bar{L}\, \mO + L\, \bO + L\bar{L}\, \kO\commae\qquad
  \mG = \mO + L\,\kO\commae
\end{equation}
\defterm{null rotation with ${\lG}$ fixed}, given by ${K\in\complexn}$,
\begin{equation}\label{lfixed}
  \kG = \kO + \bar{K}\, \mO + K\, \bO + K\bar{K}\, \lO\commae\qquad
  \lG = \lO \commae\qquad
  \mG = \mO + K\,\lO\commae
\end{equation}
\defterm{boost in the ${\kG\textdash\lG}$ plane}, ${B\in\realn}$, and
a \defterm{spatial rotation in the ${\mG\textdash\bG}$ plane}, ${\intTphi\in\realn}$,
\begin{equation}\label{boostrotation}
  \kG = B\,\kO \comma\qquad
  \lG = B^{-1}\, \lO \comma\qquad
  \mG = \exp(i\intTphi)\,\mO \period
\end{equation}
The transformations of the corresponding normalized spinor frames are listed in
appendix~\ref{apx:spinors},  relations \eqref{kfixedspinor}-\eqref{boostrotspinor}.

\subsection{The tetrad adjusted to $\scri$}
\label{ssc:AdjTetr}

\begin{figure}
\begin{center}
\includegraphics[scale=0.75]{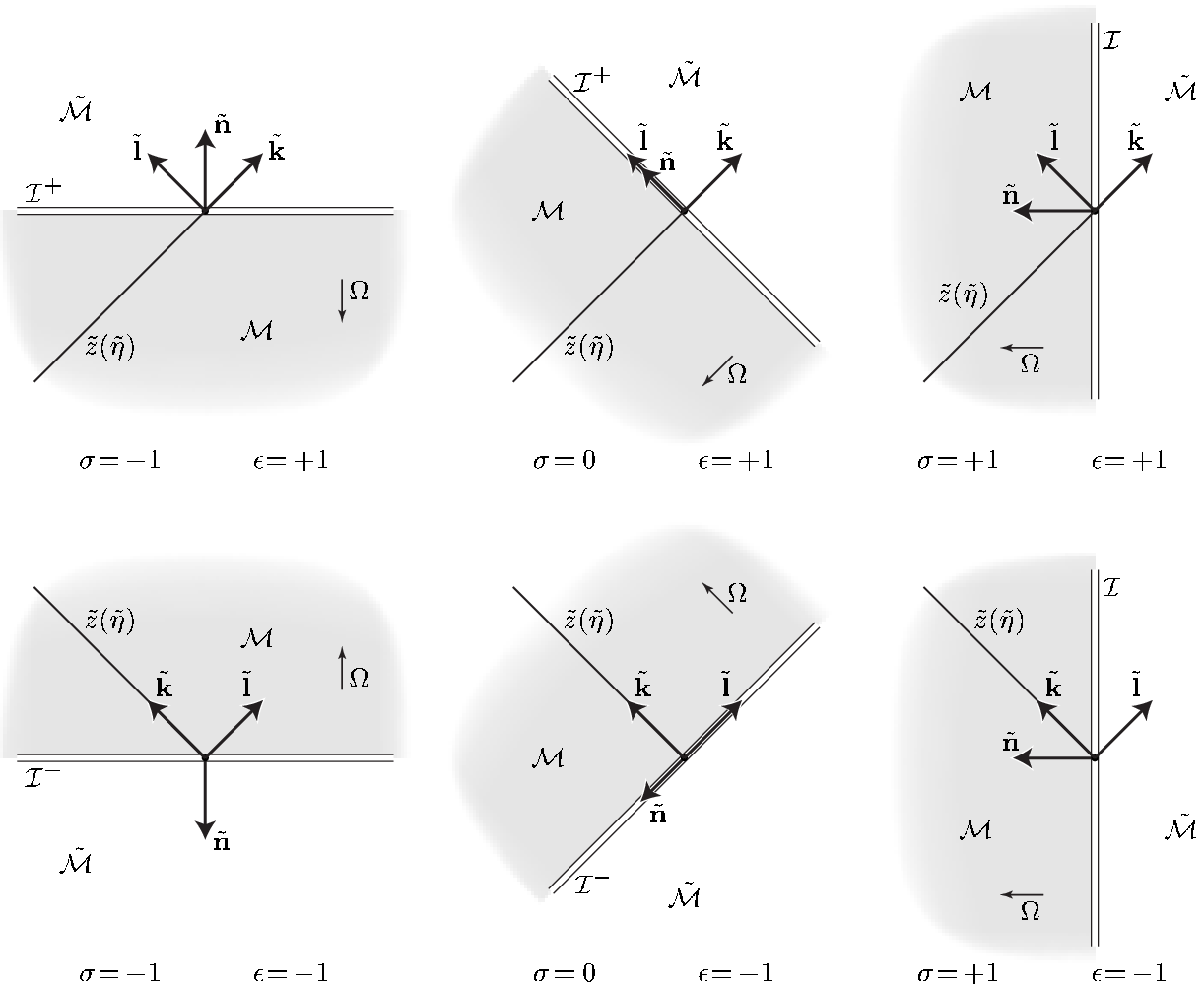}
\end{center}
\caption{\label{fig:adjtetr}%
Tetrads adjusted to conformal infinity $\scri$ of various character, determined by $\nsgn$,
cf. figure~\ref{fig:scri}. If the vector $\kC$ is oriented along the tangent vector of
the null geodesic $\cgeod(\cafp)$ it is either outgoing (${\EPS=+1}$) or ingoing (${\EPS=-1}$).}
\end{figure}

We say that a conformal null tetrad is \defterm{adjusted to conformal infinity\,}
if the vectors $\kC$ and $\lC$  on $\scri\,$ are coplanar with $\cnorm$ (the vector
\eqref{NormVect} normal to conformal infinity), and satisfy the relation
\begin{equation}\label{Adjusted}
   \cnorm = \EPS\textstyle{\frac{1}{\sqrt{2}}}
   (-\nsgn\kC+\lC)\commae
\end{equation}
where ${\EPS=\pm1}$.
As shown in figure~\ref{fig:adjtetr}, for a \vague{de~Sitter type} spacelike infinity (${\nsgn=-1}$) there is
${\cnorm = \EPS\,\tC= \EPS\,(\kC+\lC)/\sqrt{2}}$, for an  \vague{anti--de~Sitter type} timelike $\scri\,$ (${\nsgn=+1}$)
 ${\cnorm = -\EPS\,\qC= -\EPS\,(\kC-\lC)/\sqrt{2}}$, and ${\cnorm = \EPS\,\lC/\sqrt{2}\,}$
for null \vague{Minkowskian} $\scri\,$ (${\nsgn=0}$).
If the null vector $\kC$ is chosen to be oriented along the
tangent vector of the null geodesic $\cgeod(\cafp)$, the
parameter $\EPS$ then identifies whether the geodesic is outgoing (${\EPS=+1}$)
or ingoing (${\EPS=-1}$), see~\eqref{DefinitionEpsilon}.
Notice that the condition \eqref{Adjusted} also implies ${\cmtrc_{ab}\mC^a\cnorm^b =
0=\cmtrc_{ab}\bC^a\cnorm^b}$, so that the vectors ${\mC,\bC}$ of the tetrad adjusted to
conformal infinity are always \emph{tangent to} $\scri$.

Analogously we  define a tetrad in the physical spacetime adjusted to conformal infinity
by the condition
\begin{equation}\label{AdjustedPhys}
   \norm = \EPS\textstyle{\frac{1}{\sqrt{2}}}
   (-\nsgn\kG+\lG)\commae
\end{equation}
where the vector $\norm$ normal to $\scri$ in ${(\mfld,\mtrc)}$ is normalized by \eqref{NormNorm}
(cf. discussion  at the end of section \ref{ssc:ConfInf}).

\subsection{The interpretation tetrad}
\label{ssc:IntTetr}
Let us  introduce an \defterm{interpretation} null
tetrad ${\kI,\,\lI,\,\mI,\,\bI}$. It is any tetrad which is \emph{parallelly
transported along a null geodesic} $\geod(\afp)$ \emph{in the physical spacetime} $\mfld$,
with $\kI$ tangent to $\geod(\afp)$.
We thus require
\begin{equation}\label{PhysK}
  \kI=\frac{\gamma}{\sqrt2\scale}\frac{D\geod}{d\,\afp} \comma
  \gamma=\text{constant}\commae
\end{equation}
and
\begin{equation}\label{PhysParallel}
  \kI^a\covd_a\kI^b=0 \comma
  \kI^a\covd_a\lI^b=0 \comma
  \kI^a\covd_a\mI^b=0 \comma
  \kI^a\covd_a\bI^b=0 \period
\end{equation}
Here, ${\scale}$ is a constant scale parameter introduced below equation \eqref{conflapse}.
For geodesic the first equation in \eqref{PhysParallel} is satisfied automatically. It only remains
to investigate the remaining three conditions for parallel transport of the
vectors ${\lI,\,\mI}$, and $\bI$.

The interpretation tetrad is not unique --- there is a freedom
in its particular initial or final specification.
It is possible to scale the vector ${\kI}$ by fixing the constant ${\gamma}$ in \eqref{PhysK}.
By such a choice we fix the \vague{physical wavelength} of
the associated null ray. The initial scale of the
vector ${\kI}$ can be fixed somewhere in the spacetime, e.g., with respect to
a Killing vector or with respect to worldlines of sources, etc.
Unfortunately, on a general level, we do not know how to specify privileged
initial conditions for the interpretation tetrad. However, our goal here is to \emph{compare}
the field measured in interpretation tetrads transported along \emph{different}
null geodesics approaching the same point at ${\scri}$.
It is thus natural to choose the \emph{final} conditions for tetrads
in a \vague{comparable} way independently of the  geodesics.
Observing that the normalization of the tangent vector ${D\geod/d\afp}$ was
chosen naturally with respect to the asymptotic structure of the spacetime
by equations \eqref{RelatAfpCafp} and \eqref{InOutGeod},
we require that the vector ${\kI}$ is proportional to ${D\geod/d\afp}$
by the \emph{same} factor. Namely, we require that the constant~${\gamma}$
in \eqref{PhysK} is independent of the choice of the geodesics.
In the following we will set
\begin{equation}\label{kIfixing}
{\gamma=1}\period
\end{equation}
This is equivalent to the condition that the component of the vector ${\kI}$ normal to
the conformal infinity is the same for \emph{all} interpretation tetrads approaching
a given point on ${\scri}$, cf.\ equation \eqref{kAscaling} below.

There is a remaining freedom in the choice of the interpretation tetrad which
corresponds to a null rotation with $\kI$ fixed,
and to a spatial ${\mI\textdash\bI}$ rotation.
However, we will find that the asymptotic characterization of the field components
derived in section~\ref{ssc:dirstrrad} does not, in fact,  depend on such freedom.
To demonstrate this property, we now analyze an explicit relation between the
interpretation tetrad and a conformal tetrad adjusted to $\scri$.

\subsection{Asymptotic behaviour of the interpretation tetrad}
\label{ssc:AsympIntTetr}
We consider a particular null tetrad ${\kP,\,\lP,\,\mP,\,\bP}$,
where $\kP$  is \emph{tangent to a null geodesic} $\cgeod(\cafp)$,
\begin{equation}\label{ConfK}
  \kP=\frac{1}{\sqrt2\scale}\frac{D\cgeod}{d\,\cafp} \commae
\end{equation}
$\lP$ is \emph{coplanar} with $\kP$ and $\cnorm$ on ${\scri}$, and
such that the vectors of the tetrad are
\emph{parallelly transported} along $\cgeod(\cafp)$
in conformal geometry,
\begin{equation}\label{ConfParallel}
  \kP^a\ccovd_a\kP^b=0 \comma
  \kP^a\ccovd_a\lP^b=0 \comma
  \kP^a\ccovd_a\mP^b=0 \comma
  \kP^a\ccovd_a\bP^b=0 \period
\end{equation}
Using the expressions \eqref{ConfK}, \eqref{InOutGeod}, and \eqref{NormVect} we
immediately derive the relation ${-\nsgn\kP+\lP=\EPS\sqrt{2}\,\cnorm}$
on conformal infinity,
which demonstrates that the tetrad considered \emph{becomes adjusted to $\scri$}\  for
${\cafp=0}$ at the point $\cgeod(0)=P\in\scri$.

On the other hand, from \eqref{PhysK}, \eqref{ConfK}, \eqref{RelatAfpCafp} we
obtain
\begin{equation}\label{KinterpKadjustRel}
\kI=\om^2\,\kP\commae
\end{equation}
 so that using \eqref{NormCnormRel},
${\mtrc_{ab}\kI^a\norm^b=-\EPS{\textstyle{\frac{1}{\sqrt{2}}}}\om}$.
The interpretation tetrad is thus \emph{not} adjusted to $\scri$ since it
does not satisfy equation \eqref{AdjustedPhys}.

To find an explicit relation between the tetrads ${\kI,\,\lI,\,\mI,\,\bI}$
and ${\kP,\,\lP,\,\mP,\,\bP}$, we first introduce an \defterm{auxiliary null tetrad}
${\kA,\,\lA,\,\mA,\,\bA}$ by conformal rescaling of the tetrad ${\kP,\,\lP,\,\mP,\,\bP}$,
\begin{equation}\label{Auxiliary}
  \kA=\om\,\kP \comma
  \lA=\om\,\lP \comma
  \mA=\om\,\mP \comma
  \bA=\om\,\bP \period
\end{equation}
The auxiliary tetrad is normalized with respect to the physical metric $\mtrc$, it is
adjusted to $\scri$ but its vectors are no longer parellelly transported along the
geodesics $\geod(\eta)$ in~$\mfld$.

Secondly, we perform a specific boost of the interpretation
tetrad such that the boost parameter is given by the conformal factor,
introducing thus the tetrad ${\kB,\,\lB,\,\mB,\,\bB}$,
\begin{equation}\label{ScaledTetrad}
  \kB=\om^{-1}\,\kI \comma
  \lB=\om\,\lI \comma
  \mB=\mI \comma
  \bB=\bI \period
\end{equation}
The vector $\kB$ is then normalized on $\scri$ in the same way as $\kA$, namely
\begin{equation}\label{kAscaling}
\mtrc_{ab}\kB^a\norm^b=-\EPS{\textstyle{\frac{1}{\sqrt{2}}}}\period
\end{equation}

Considering \eqref{KinterpKadjustRel}, the tetrad \eqref{ScaledTetrad}
has to be related to the auxiliary tetrad \eqref{Auxiliary} by a
null rotation  with fixed $\kG$, and a possible spatial rotation in the ${\mG\textdash\bG}$ plane,
\begin{equation}\label{NullRotAux}
\begin{aligned}
  \kB &= \kA  \commae\\
  \lB &= \lA+\bar{\bstTL}\,\mA+\bstTL\,\bA+\bstTL\bar{\bstTL}\,\kA  \commae\\
  \mB &= \exp(i\intTphi)(\mA+\bstTL\,\kA)  \commae
\end{aligned}
\end{equation}
with parameters ${\bstTL\in\complexn}$, ${\intTphi\in\realn}$, cf.\ \eqref{kfixed}, \eqref{boostrotation}.
For the interpretation tetrad it implies
\begin{equation}\label{NullRotAuxExpl}
\begin{aligned}
  \kI &= \om\,\kA  \commae\\
  \lI &= \om^{-1}\lA+\bar\intTL\,\mA+\intTL\,\bA+\intTL\bar\intTL\,\om\,\kA  \commae\\
  \mI &= \exp(i\intTphi)(\mA+\intTL\,\om\,\kA)  \commae
\end{aligned}
\end{equation}
with ${\intTL=\om^{-1}\bstTL}$.
Now, substituting these expressions into the conditions \eqref{PhysParallel} for parallel
transport of the interpretation tetrad, and using \eqref{DerivRelation}, \eqref{ConfParallel}
and \eqref{KinterpKadjustRel} we obtain
\begin{equation}\label{RelConfInterp}
  \kP^a \,\grad_a \intTphi = 0  \comma\qquad
  \kP^a \,\grad_a \intTL    = \om^{-2}\,\mP^a \,\grad_a\om  \period
\end{equation}
The first equation implies ${\intTphi=\intTphi_{0}=\text{const.}}$, see equation \eqref{ConfK}.
Assuming again the regularity of conformal geometry near infinity, the term on
the right-hand side of the second equation can be expanded in powers of $\cafp$. Moreover, for ${\cafp=0}$
the vector $\mP$ is tangent to  $\scri$, see the text below equation
\eqref{Adjusted}, so that the expansion has the form
\begin{equation}\label{Mepansion}
  {\sqrt2\,\scale}\,\mP^a
  \,\grad_a\om=M_1\,\cafp+M_2\,\cafp^2+\ldots\commae
\end{equation}
where ${M_1, M_2}$ are constants which depend on derivatives of $\om$.
Using \eqref{OmegaAnalyt} we thus integrate \eqref{RelConfInterp} to get
\begin{equation}\label{LinCafp}
   \intTL = M_{1}\ln|\cafp|+\intTL_{0} +\ldots \comma\quad\text{i.e.,}\quad
   \bstTL  = -\EPS M_{1}\,\cafp\ln|\cafp|-\EPS\intTL_{0}\,\cafp+\ldots\commae
\end{equation}
where ${\intTL_{0}}$ is a constant of integration. Using \eqref{CafpOfAfp}
we  obtain the expansion in physical affine parameter $\afp$,
\begin{equation}\label{PhiLinAfp}
\begin{aligned}
   \intTphi &= \intTphi_0\commae\\
   \intTL &= - M_{1}\ln|\afp|+\intTL_{0} +\ldots \comma\quad\text{i.e.,}\quad
   \bstTL  = -\EPS M_{1}\,\afp^{-1}\ln|\afp|+\EPS\intTL_{0}\,\afp^{-1}+\ldots\period
\end{aligned}
\end{equation}
Calculations to the next order in the affine parameter are presented in appendix~\ref{apx:expansions}.

We observe that ${\bstTL}$ \emph{approaches zero near} $\scri$.
Inspecting the relations \eqref{NullRotAux} we thus obtain an important result:
\emph{the tetrad ${\kB,\,\lB,\,\mB,\,\bB}$}, associated with \emph{any
interpretation tetrad} by the boost \eqref{ScaledTetrad},
\emph{becomes asymptotically adjusted to conformal infinity\, $\scri$}.
Asymptotically it may differ from the chosen auxiliary tetrad ${\kA,\,\lA,\,\mA,\,\bA}$ only by a trivial
rotation in ${\mA\textdash\bA}$ plane by a fixed angle ${\intTphi_{0}}$,
and in this sense it is \vague{essentially unique}.

Let us note,  that in general the vectors ${\kI,\,\lI}$ of the interpretation tetrad are not asymptotically
coplanar with the normal ${\norm}$ to $\scri$. From equations \eqref{NullRotAuxExpl}
with ${\intTL}$ given by \eqref{LinCafp} we see that the vector ${\lI}$ has components
in the ${\mA\textdash\bA}$ directions which are perpendicular to ${\norm}$.
Fortunately, these components grow only logarithmically and, as we shall see later,
they do not influence the leading term of the fields evaluated with respect to the
interpretation tetrad. Moreover, it is demonstrated in appendix~\ref{apx:expansions} (see equation
\eqref{LogTermMissing3}) that
\begin{equation}\label{LogTermMissing3A}
M_1
\sim \bigl(\cafp^{-1}\,\RicciP{\auxT}{01} \bigr)\big|_{\cafp=0}
\sim \bigl(\afp\,\RicciP{\auxT}{01} \bigr)\big|_{\afp=\EPS\infty}  \commae
\end{equation}
where ${\RicciP{\auxT}{01}}$ is the specific component of the energy-momentum tensor evaluated
in the auxiliary tetrad \eqref{Auxiliary}. This vanishes  for
a vacuum spacetime. It also disappears in non-vacuum cases when the asymptotic Einstein
condition \eqref{AsymEinstCond} holds --- it is satisfied  provided that  matter
fields decay  faster then ${\,\sim\cafp\,}$ near conformal infinity.
For such spacetimes, the ${\,\ln|\cafp|\,}$ term in the expansion \eqref{LinCafp} of the
parameter $\intTL$ near $\scri$ is absent, so that
\begin{equation}\label{LExplicVacA}
\intTL \lteq \intTL_0\comma\quad\text{i.e.,}\quad
\bstTL \lteq -\EPS \intTL_0\,\cafp\commae
\end{equation}
and, by the proper choice ${\intTL_0=0}$, the vectors ${\kI,\,\lI}$ of  the
interpretation tetrad \emph{can be arranged} to become
asymptotically \emph{coplanar} with the normal ${\norm}$.
This coplanarity was assumed previously in \cite{PenroseRindler:book}
(see discussion concerning figure 9-20 therein), and in
\cite{KrtousPodolskyBicak:2003,KrtousPodolsky:2004a}.

Let us also discuss the geometrical meaning of the integration constants
${\intTL_0}$ and ${\intTphi_0}$. In the above derivation we have represented
the transformation \eqref{NullRotAuxExpl} from the auxiliary tetrad
to the interpretation tetrad as an application of null rotation
with fixed ${\kG}$ given by the parameter ${\bstTL=\bstTL_* + \om \intTL_0}$
($\bstTL_*$ independent of ${\intTL_0}$, cf.\ equation~\eqref{LinCafp}),
spatial rotation with the parameter ${\intTphi=\intTphi_0}$,
and finally boost with the parameter ${B=\om}$. This can be rearranged as
the sequence of ${\kG}$-fixed null rotation given by the parameter ${\bstTL_*}$,
boost with ${B=\om}$, then ${\kG}$-fixed null rotation given by the parameter
${\intTL_0}$, and finally spatial rotation with ${\intTphi=\intTphi_0}$.
The last two transformations exactly correspond to the freedom in
the choice of the interpretation tetrad --- the condition \eqref{PhysK}
determines the interpretation tetrad exactly up to such null rotation with ${\kG}$
fixed, and a spatial rotation in the ${\mG\textdash\bG}$ plane.
The parameters ${\intTL_0}$ and ${\intTphi_0}$ thus determine
a \emph{specific choice} of the interpretation tetrad which is usually given by some
physical prescription in a finite domain of the spacetime.

It will be demonstrated below that the \emph{asymptotic directional
behaviour of the fields} (see section~\ref{ssc:fieldinint})
\emph{is independent of the parameter} ${\intTL_0}$.
It will depend on the parameter ${\intTphi_0}$ only through
a \emph{phase} of the complex component of the field, and such dependence can
be eliminated by considering just the magnitude of the field.
We will return to the corresponding question of the phase (\vague{polarization})
dependence of the fields in the discussion of the results.

\subsection{The reference tetrad and parameterization of null directions}
\label{ssc:RefTetr}

In the following, we will need to identify the
direction ${\kI}$ of the null geodesic and orientation of the associated interpretation tetrad
near conformal infinity using suitable directional parameters.
For this purpose we set up a reference tetrad. The \defterm{reference tetrad} ${\kO,\,\lO,\,\mO,\,\bO}$ is
any tetrad adjusted to $\scri$ which satisfies the
coplanarity and normalization condition \eqref{AdjustedPhys},
\begin{equation}\label{AdjustedRefer}
   \norm = \EPS_\refT\textstyle{\frac{1}{\sqrt{2}}}(-\nsgn\kO+\lO)\period
\end{equation}
Otherwise, the reference tetrad can be chosen \emph{arbitrarily}, ergo conveniently.
It may thus either respect the symmetry of the spacetime (by adapting the
reference tetrad to the Killing vectors) or its specific algebraic
structure (in which case it can be oriented along the principal
null directions). The parameter ${\EPS_\refT=\pm1}$ in \eqref{AdjustedRefer}
will be chosen in such a way that the vectors ${\kO}$ and ${\lO}$
are \emph{future oriented}. For ${\nsgn=-1,\,0}$ this means that ${\EPS_\refT=+1}$
on ${\scri^+}$, and ${\EPS_\refT=-1}$ on ${\scri^-}$.
For ${\nsgn=+1}$ the parameter $\EPS_\refT$  can be chosen either $+1$ or $-1$: it
corresponds to $\kO$ oriented outside or inside $\mfld$, cf.\ figure~\ref{fig:adjtetr}.

We use the given reference tetrad ${\kO,\,\lO,\,\mO,\,\bO}$ as a
fixed basis with respect to which it is possible to define uniquely other
directions, for example asymptotic directions along which
various null geodesics approach a point $P$ at $\scri$,
or the principal null directions, see section~\ref{ssc:PND}.
It is natural to characterize such a general
null direction $\kG$ by a \emph{complex parameter} ${\R}$ in the following
way: the direction $\kG$ is obtained (up to a rescaling) from $\kO$ by the null
rotation \eqref{lfixed} with the parameter ${K=R}$,
\begin{equation}\label{Rmeaning}
  \kG \propto\kO + \bar{\R}\, \mO + \R\, \bO + \R\bar{\R}\, \lO    \period
\end{equation}
The value ${\R=\infty}$ is also permitted --- this corresponds to
$\kG$ being oriented along $\lO$.

Let us mention that the reference tetrad introduced above is not well defined in
conformal geometry --- it is normalized using the physical metric.
However, it could be rescaled \emph{isotropically} by the \emph{common} factor ${\om^{-1}}$
to obtain the associated \defterm{conformal reference tetrad} which is well defined in the conformal
geometry. For convenience, in the following we will use the physically normalized
reference tetrad instead of the conformally normalized one ---
see  discussion at the end of section \ref{ssc:ConfInf}.

\section{The fields and their asymptotic structure}
\label{sc:fields}

Now we have all \vague{prerequisites} needed to analyze the asymptotic properties of the fields.
We are mainly interested in gravitational and
electromagnetic fields. However,  the principal result
--- the asymptotic directional structure of the fields ---
can be derived for general fields of spin ${s}$.
In all these cases we will study the dominant (radiative) component of the field
as one approaches conformal infinity. For this purpose, it is useful to
parametrize the fields using complex tetrad components which have  special
transformation properties.

\subsection{The field components and their transformation properties}
\label{ssc:fieldtrans}
Following the notation of \cite{Stephanietal:book}, gravitational field is characterized
by the Weyl tensor $\WT_{abcd}$ and can be parametrized by five complex coefficients
\begin{equation}\label{PsiDef}
\begin{aligned}
  \WTP{}{0} &=  \WT_{abcd}\, \kG^a \,\mG^b \,\kG^c \,\mG^d \comma\\
  \WTP{}{1} &=  \WT_{abcd}\, \kG^a \,\lG^b \,\kG^c \,\mG^d \commae\\
  \WTP{}{2} &=  \WT_{abcd}\, \kG^a \,\mG^b \,\bG^c \,\lG^d \commae\\
  \WTP{}{3} &=  \WT_{abcd}\, \lG^a \,\kG^b \,\lG^c \,\bG^d \comma\\
  \WTP{}{4} &=  \WT_{abcd}\, \lG^a \,\bG^b \,\lG^c \,\bG^d \commae
\end{aligned}
\end{equation}
whereas electromagnetic field is described by the tensor $\EMF_{ab}$ which is
parametrized by three complex coefficients
\begin{equation}\label{PhiDef}
\begin{aligned}
  \EMP{}{0} &= \EMF_{ab}\, \kG^a \,\mG^b \comma\\
  \EMP{}{1} &= {\textstyle\frac12}\,\EMF_{ab}\,
        \bigl(\kG^a \,\lG^b -\mG^a \,\bG^b \bigr)\commae\\
  \EMP{}{2} &= \EMF_{ab}\, \bG^a \,\lG^b \period
\end{aligned}
\end{equation}
By a field of spin $s$ we understand field which transforms
according to spin-${s}$ representation of the Lorentz group.
It can be characterized by ${(2s+1)}$ complex components
\begin{equation}\label{GenField}
\ \fieldP{}{j}\comma\quad j=0,1,\ldots,2s\period
\end{equation}
A more detailed (spinor) description of such fields can be found in appendix~\ref{apx:spinors}.
The gravitational field $\WTP{}{j}$ or electromagnetic field $\EMP{}{j}$ are special
cases of \eqref{GenField} for ${s=2,1}$, cf.~\eqref{Wspinor}-\eqref{spincompon}.

These field components transform  in a well-known way under special Lorentz
transformations introduced above  (see, e.g., Ref.~\cite{Stephanietal:book} or
appendix~\ref{apx:spinors}). For a null rotation with $\kG$ fixed, cf.\ equation~\eqref{kfixed},
the field transforms as
\begin{equation}\label{kfixedField}
 \fieldP{}{j} = \fieldP{\refT}{j}+j\bar{L}\,\fieldP{\refT}{j-1}
  +\binom{j}{2}\bar{L}^2\,\fieldP{\refT}{j-2}+\binom{j}{3}\bar{L}^3\,\fieldP{\refT}{j-3}
  +\ldots+\bar{L}^j\,\fieldP{\refT}{0}  \period
\end{equation}
Under a null rotation with $\lG$ fixed, see equation~\eqref{lfixed},
the transformation reads
\begin{equation}\label{lfixedField}
 \fieldP{}{j} = \fieldP{\refT}{j}+(2s-j)K\,\fieldP{\refT}{j+1}
  +\binom{2s-j}{2}K^2\,\fieldP{\refT}{j+2}+\ldots+K^{2s-j}\,\fieldP{\refT}{2\!s}  \period
\end{equation}
Under a boost in the ${\kG\textdash\lG}$ plane, and a spatial rotation in the ${\mG\textdash\bG}$ plane,
given by equation~\eqref{boostrotation},
the components $\fieldP{}{j}$ transform as
\begin{equation}\label{boostrotationField}
 \fieldP{}{j} = B^{s-j}\,\exp\bigl(i(s-j)\intTphi\bigr)\; \fieldP{\refT}{j} \period
\end{equation}

\subsection{Principal null directions and algebraic classification}
\label{ssc:PND}
For gravitational,  electromagnetic, or any  spin-$s$ field there exist
\defterm{principal null directions}  (PNDs) which  are
privileged null directions $\kG$ such that ${ \fieldP{}{0}=0}$
in a null tetrad ${\kG,\,\lG,\,\mG,\,\bG}$ (a choice of ${\lG,\,\mG,\,\bG}$ is irrelevant).
The PND $\kG$ can be obtained from a reference tetrad ${\kO,\,\lO,\,\mO,\,\bO}$
by a null rotation~\eqref{Rmeaning} given by a directional parameter ${\R\in\complexn}$.
We choose the remaining vectors ${\lG,\,\mG,\,\bG}$ to be given by the same null rotation,
i.e., by \eqref{lfixed} with ${K=\R}$.
The condition ${\fieldP{}{0}=0}$ thus takes the form
of an algebraic equation of the order $2s$ for the directional parameter $R$, see
equation~\eqref{lfixedField},
\begin{equation}\label{rootsPND}
\R^{2\!s}\,\fieldP{\refT}{2\!s}+\binom{2s}{2s\!-\!1}\,\R^{2\!s-1}\,\fieldP{\refT}{2\!s-1}
   +\ldots + \binom{2s}{1}\,\R\,\fieldP{\refT}{1}+\fieldP{\refT}{0} =0\period
\end{equation}
In particular, for gravitational field it reduces to a quartic
\begin{equation}\label{4rootsPND}
\R^4\, \WTP{\refT}{4} + 4 \R^3\, \WTP{\refT}{3} +
    6 \R^2\, \WTP{\refT}{2} + 4 \R\, \WTP{\refT}{1} + \WTP{\refT}{0}=0\period
\end{equation}
The complex roots $\R_n$, ${\,n=1,2,\ldots,2s}$, of equation \eqref{rootsPND}
parametrize PNDs $\kG_n$  with respect to the reference tetrad
${\kO,\,\lO,\,\mO,\,\bO}$. The situation when ${\fieldP{\refT}{2\!s}=0}$
formally corresponds to an infinite value of one of the roots, say
${\R_1=\infty}$, in which case ${\kG_1=\lO}$.
There are 4 principal null directions for gravitational field,
2 for eletromagnetic field, and $2s$ for spin-$s$ field. According to whether some of
these PNDs coincide, the fields are algebraically special and can be classified
to various (Petrov) algebraic types \cite{Penrose:1960,Stephanietal:book,PenroseRindler:book}.

In a generic situation   ${\fieldP{\refT}{2\!s}}$ is non-vanishing, and the polynomial
on the left-hand side of equation \eqref{rootsPND} can be decomposed as
\begin{gather}
\begin{aligned}
\R^{2\!s}\,\fieldP{\refT}{2\!s}+\binom{2s}{1}\,\R^{2\!s-1}\,\fieldP{\refT}{2\!s-1}+&\ldots
   + \binom{2s}{2s\!-\!1}\,\R\,\fieldP{\refT}{1}+\fieldP{\refT}{0}\\
   &=\fieldP{\refT}{2\!s}\,(\R-\R_1)(\R-\R_2)\ldots(\R-\R_{2\!s}) \period
\end{aligned}\label{DecompRootsPND}
\end{gather}
By comparing the coefficients of various powers of $\R$ it is  possible to express
all $\fieldP{\refT}{j}$ components in terms of $\fieldP{\refT}{2\!s}$ and
the algebraically privileged principal null directions characterized by $\R_n$.
For example, the components of gravitational field can be written
\begin{gather}
\begin{aligned}
  \WTP{\refT}{3} &= -{\textstyle\frac14}\,\WTP{\refT}{4}\;(\R_1+\R_2+\R_3+\R_4)\commae\\
  \WTP{\refT}{2} &= \spcm{\textstyle\frac16}\,\WTP{\refT}{4}\;(\R_1 \R_2+\R_1 \R_3+\R_1 \R_4+
                  \R_2 \R_3+\R_2 \R_4+\R_3 \R_4)\commae\\
  \WTP{\refT}{1} &= -{\textstyle\frac14}\,\WTP{\refT}{4}\;(\R_1 \R_2 \R_3+\R_1 \R_2 \R_4+
                  \R_1 \R_3 \R_4+\R_2 \R_3 \R_4)\commae\\
  \WTP{\refT}{0} &= \spcm\spcm\WTP{\refT}{4}\; \R_1 \R_2 \R_3 \R_4\period
\end{aligned}\label{GFieldinPND}
\end{gather}
Similar expressions apply to other fields; we write only the expression for the $\fieldP{\refT}{0}$ component,
\begin{equation}\label{zeroComp}
\fieldP{\refT}{0}=(-1)^{2\!s}\,\fieldP{\refT}{2\!s}\prod_{j=1}^{2\!s}\R_j \period
\end{equation}

Finally, let us note that a rescaling of all field components by a common
factor does not change the algebraic structure, i.e., the PNDs remain unchanged.
This observation is useful when we study the algebraic structure of the fields near
conformal infinity. As we will discuss in section \ref{ssc:fieldinref}, the
field components ${\fieldP{\refT}{j}}$ decay to zero near  conformal infinity,
see equation \eqref{fieldsnearscri}.
However, the leading term of the field still carries information about its
algebraic structure. In other words, asymptotically we define PNDs
in terms of the leading order of the field.

\subsection{Field components in the interpreation tetrad}
\label{ssc:fieldinint}
Using the above quantities and relations we may now analyze the asymptotic behaviour of a general
gravitational, electromagnetic or any spin-$s$ field with respect to the interpretation tetrad
near conformal infinity. To evaluate the field components $\fieldP{\intT}{j}$
in the interpretation tetrad ${\kI,\,\lI,\,\mI,\,\bI}$ we employ its relation to the
tetrad ${\kB,\,\lB,\,\mB,\,\bB}$ which is asymptotically adjusted  to $\scri$, then the relation
between this tetrad and the auxiliary tetrad ${\kA,\,\lA,\,\mA,\,\bA}$, and finally we
perform the transformation to the reference tetrad ${\kO,\,\lO,\,\mO,\,\bO}$. We thus express
$\fieldP{\intT}{j}$ in terms of $\fieldP{\refT}{j}$.

The tetrads ${\kI,\,\lI,\,\mI,\,\bI}$ and ${\kB,\,\lB,\,\mB,\,\bB}$
are related by the boost \eqref{ScaledTetrad}, i.e.,
${\kI=\om\,\kB}$, ${\lI=\om^{-1}\,\lB}$, where ${\,\om=B\lteq\EPS\,\afp^{-1}}$,
see equation \eqref{OmegaInAfp}. The next transformation to
the auxiliary tetrad is given by the ${\lG}$-fixed null rotation and spatial
rotation \eqref{NullRotAux}, with the parameters ${\bstTL}$ and ${\intTphi}$ given by
\eqref{PhiLinAfp}. As we have already demonstrated above, with ${\intTphi_0=0}$ these two
tetrads asymptotically  coincide.
Using \eqref{boostrotationField} and \eqref{kfixedField}  we thus obtain
\begin{equation}\label{IntRelParField}
 \fieldP{\intT}{j} \lteq ({\EPS\,\afp})^{j-s}\, \exp\bigl(i(s-j)\intTphi_0\bigr)\, \fieldP{\auxT}{j} \period
\end{equation}
We observe that the field components are asymptotically independent
of the parameter ${\intTL_0}$, and they depend on the parameter ${\intTphi_0}$ only through
the phase. Because these parameters ${\intTL_0}$ and ${\intTphi_0}$ specify the choice of
the interpretation tetrad, we have thus explicitly demonstrated that the
\emph{magnitude of the leading term of field components is independent of
a particular choice of the interpretation tetrad}.

The specific phase behaviour of $\fieldP{\intT}{j}$ under spatial rotations indicates
that different field components have different polarization properties.
The polarization can carry important physical data. However, to
retrieve such information it would be ne\-cessary to fix the initial conditions for the
interpretation tetrad somewhere in a finite domain of the physical spacetime.
In the general situation which we study here, we are not able to
fix the interpretation tetrad in such a complete way, and thus the
polarization information contained in the phase of the field components is not
accessible. Therefore, in the following we will concentrate on the magnitude of
the field components, and for simplicity we choose ${\intTphi_0=0}$.

Considering that all ${\fieldP{\auxT}{j}}$ for ${j=0,1,\ldots,2s}$ are of the same order, cf.\ equation \eqref{fieldsnearscri},
the expression \eqref{IntRelParField} demonstrates the well-known
\defterm{peeling-off property} of the fields according to which various
tetrad components are proportional to different powers of the
affine parameter $\afp$ as one approaches conformal infinity along a null
geodesic. The \emph{dominant component} is ${j=2s}$.  Such a term ${\fieldP{\intT}{2\!s}}$
represents the radiative part of the field. In particular, the
dominant component of the gravitational field is characterized by
${\WTP{\intT}{4} \lteq {\afp}^{2}\, \WTP{\auxT}{4}}$, the
electromagnetic field by ${\EMP{\intT}{2} \lteq \EPS\afp\, \EMP{\auxT}{4}}$, etc.

Finally, we express ${\fieldP{\auxT}{2\!s}}$ in terms of components $\fieldP{\refT}{j}$. Both the reference
and the auxiliary tetrads are adjusted to ${\scri}$ and thus they only
differ by a transformation which leaves the normal vector ${\norm}$ fixed. Such a transformation
can be obtained, e.g., by the null rotation \eqref{kfixed} of the reference tetrad,
followed by the null rotation \eqref{lfixed}, and the boost \eqref{boostrotation}
with the parameters
\begin{gather}
\begin{aligned}
  L &= \nsgn \R\commae\\
  K &= \frac{\R}{1-\nsgn \R\bar{\R}}\commae\\
  B &= \EPS\EPS_\refT({1-\nsgn \R\bar{\R}})\period
\end{aligned}\label{ParamRefAux}
\end{gather}
(For a general transformation between two tetrads adjusted to
${\scri}$ we should also admit a spatial rotation but this only
changes a phase of the field components which was discussed above.)
It has an explicit form
\begin{gather}
\begin{aligned}
  \kA &= \frac{\EPS\EPS_\refT}{1-\nsgn\R\bar{\R}}\biggl({\kO + \bar{\R}\, \mO + \R\, \bO + \R\bar{\R}\, \lO}\biggr)  \commae\\
  \lA &= \frac{\EPS\EPS_\refT}{1-\nsgn\R\bar{\R}}\biggl({\nsgn^2\R\bar{\R}\, \kO + \nsgn\bar{\R}\, \mO + \nsgn\R\, \bO + \lO}\biggr)\commae\\
  \mA &= \frac{1}{1-\nsgn\R\bar{R}}\biggl({\nsgn\R\,\kO + \mO + \nsgn\R^2\, \bO + \R\, \lO}\biggr)   \period
\end{aligned}\label{kAndlAux}
\end{gather}
Using \eqref{Auxiliary} and \eqref{AdjustedRefer} we  easily check that
${\EPS\textstyle{\frac{1}{\sqrt{2}}}(-\nsgn\kP+\lP)=\cnorm}$,
which is the condition \eqref{Adjusted}.
Moreover, the vector $\kA$ satisfies \eqref{Rmeaning},
and represents the direction along which the null geodesic
approaches conformal infinity: this direction is characterized by
the complex directional parameter $\R$.

It only remains to perform the transformation of the leading field component
corresponding to \eqref{ParamRefAux}. Using  \eqref{kfixedField}
--\eqref{boostrotationField} we obtain
\begin{equation}\label{TransfFieldtoAux}
\fieldP{\auxT}{2\!s} = B^{-s}\bar{L}^{2\!s}\,\biggl[\bar{L}^{-2\!s}\,\fieldP{\refT}{2\!s}+\binom{2\!s}{1}\bar{L}^{-2\!s+1}\,\fieldP{\refT}{2\!s-1}
  +\binom{2\!s}{2}\bar{L}^{-2\!s+2}\,\fieldP{\refT}{2\!s-2} +\ldots+\fieldP{\refT}{0}\biggr]  \period
\end{equation}
Applying now the identity \eqref{DecompRootsPND},
the expression in the bracket can be written as
${\fieldP{\refT}{2\!s}\,(\bar{L}^{-1}-\R_1)(\bar{L}^{-1}-\R_2)\ldots(\bar{L}^{-1}-\R_{2\!s})}$,
so that
\begin{equation}\label{FieldInAux}
\fieldP{\auxT}{2\!s} = \fieldP{\refT}{2\!s}\,B^{-s}\,
(1-\R_1\bar{L})(1-\R_2\bar{L})\ldots(1-\R_{2\!s}\bar{L})  \period
\end{equation}
Using \eqref{IntRelParField}, \eqref{ParamRefAux} we thus obtain explicitly
\begin{equation}\label{FieldInAuxExpl}
\fieldP{\intT}{2\!s} \lteq  \fieldP{\refT}{2\!s}\, \Bigl(\frac{\EPS_\refT\,\afp}{1-\nsgn \R\bar{\R}}\Bigr)^{s}\,
\bigl(1-\nsgn\R_1\bar{\R}\bigr)\bigl(1-\nsgn\R_2\bar{\R}\bigr)\ldots\bigl(1-\nsgn\R_{2\!s}\bar{\R}\bigr)  \period
\end{equation}

\subsection{Asymptotic behaviour of the field components in the reference tetrad}
\label{ssc:fieldinref}

For a complete analysis of radiation it is important to identify the specific \vague{fall-off}
of the field. The correct asymptotic behaviour can only be obtained
by a detailed study of the field equations. There exists a wide spectrum of various
results concerning this topic in the literature. As mentioned in the introduction, the decay behaviour
of the fields is well understood in asymptotically flat spacetimes, and there
are some important results also in the case of a non-vanishing cosmological constant.

However, our goal in this work is to study the \emph{directional dependence} of the leading term of the fields,
not its \emph{decay bahaviour}. We will thus only assume
the fall-off typical for \emph{zero-rest-mass fields}, without engaging in a
study of the field equations. Motivated by discussion of behaviour of fields
with a consistent field equation (${s\leq2}$) in asymptotically flat spacetimes
(\cite{Penrose:1965,PenroseRindler:book} or, e.g., \cite{Friedrich:2002,Frauendiener:2004} for gravitational field)
we will assume
\begin{equation}\label{fieldsnearscri}
  \fieldP{\refT}{j}\lteq\frac{\fieldP{\refT}{j}{}_*}{\afp^{s+1}}\comma\qquad
  \fieldP{\refT}{j}{}_*=\text{constant}\period
\end{equation}
For gravitational and electromagnetic fields this means that
${\WTP{\refT}{j}\lteq\WTP{\refT}{j}{}_*\;\afp^{-3}}$,
${\EMP{\refT}{j}\lteq\EMP{\refT}{j}{}_*\;\afp^{-2}}$.
Recalling the behaviour \eqref{OmegaInAfp} of the conformal factor
and the fact that tensor of electromagnetic field and the Weyl tensor
are conformally invariant, ${{\tilde\EMF}_{ab}=\EMF_{ab}}$, ${{\tilde\WT}_{abc}{}^{d}=\WT_{abc}{}^{d}}$,
the fall-off \eqref{fieldsnearscri} follows from the condition that
the conformal quantities ${{\tilde\EMF}_{ab}}$ and
${\tens{d}_{abc}{}^{d}=\om^{-1}\tilde{\WT}_{abc}{}^{d}}$ are regular at infinity.
For ${\Lambda\ne0}$ such behaviour of gravitational field can be obtained rigorously, see, e.g.,
\cite{Friedrich:1998a,Friedrich:2002}, and it is plausible also for asymptotically flat spacetimes.
Inspired by these observations, in the following we will assume the behaviour \eqref{fieldsnearscri}
in a general situation.

Of course, some of the field components may decay faster
even if the fall-off \eqref{fieldsnearscri} is valid for a generic component.
This happens when the reference tetrad is aligned
along PNDs, as we will discuss in the next section. If at least one of the field
components ${\fieldP{\refT}{j}}$ falls-off as in \eqref{fieldsnearscri} (i.e., at least
one ${\fieldP{\refT}{j}{}_*}$ is non-vanishing) it is always
possible to change the reference tetrad in such a way that all
${\fieldP{\refT}{j}{}_*\neq0}$.
When all field components ${\fieldP{\refT}{j}}$ decay faster than
\eqref{fieldsnearscri} we call such a field to be asymptotically of type {0}, i.e., the field
with a trivial algebraic structure.

Let us however emphasize again that the assumption \eqref{fieldsnearscri} \emph{is not crucial} for
the  asymptotic directional structure of the field. It influences
the decay of the field, not its directional dependence.
Because we are mainly interested in the analysis of the directional structure
we will not study the behaviour \eqref{fieldsnearscri} in more detail.

\subsection{Asymptotic directional structure of radiation}
\label{ssc:dirstrrad}

Substituting  \eqref{fieldsnearscri} into equation \eqref{FieldInAuxExpl} we finally obtain
\begin{equation}\label{FieldInterpComplete}
\fieldP{\intT}{2\!s} \lteq \frac{1}{\afp} \,\,\EPS_\refT^s\fieldP{\refT}{2\!s}{}_*\,
  \frac{\bigl(1-\nsgn\R_1\bar{\R}\bigr)\bigl(1-\nsgn\R_2\bar{\R}\bigr)\ldots\bigl(1-\nsgn\R_{2\!s}\bar{\R}\bigr)}
  {\bigl(1-\nsgn \R\bar{\R}\bigr)^s}\,  \period
\end{equation}
This expression fully characterizes the asymptotic behaviour on $\scri$ of the dominant
component of any massless field of spin $s$ in the normalized interpretation tetrad
${\kI,\,\lI,\,\mI,\,\bI}$ which is parallelly propagated
along a null geodesic $\geod(\afp)$. Due to the remaining freedom corresponding to
a spatial rotation \eqref{boostrotation} in the transverse ${\mI\textdash\bI}$ plane,
only the modulus ${|\fieldP{\intT}{2\!s}|}$ has an invariant meaning, the phase
of ${\fieldP{\intT}{2\!s}}$ describes a polarization.
The field decays as ${\afp^{-1}}$, where $\afp$ is the affine parameter, so we call the
expression \eqref{FieldInterpComplete} the \emph{radiative part of the field}.

The complex parameter $\R$ represents the \emph{direction} of
the null geodesic along which a given point ${P\in\scri}$ of conformal infinity
is approached as ${\afp\to\EPS\infty}$. Let us recall that
the constants $\R_n$ characterize the \emph{principal null directions},
i.e. the algebraic structure  of the field at $P$. The directional
structure of radiation is thus completely determined by the algebraic
(Petrov) type of the field. However, the dependence of
$\fieldP{\intT}{2\!s}$ on the direction $\R$ along which $P\in\scri$
is approached occurs only if $\nsgn\not=0$, i.e., at a
\vague{de~Sitter-like} or \vague{anti--de~Sitter-like}  conformal infinity.
For $\scri$ of \vague{Minkowskian} type  which has a null
character, ${\nsgn=0}$, this directional dependence completely vanishes.

The directional pattern of radiation \eqref{FieldInterpComplete} has been
derived assuming that the field component ${\fieldP{\refT}{2s}}$ is non-vanishing,
cf.\ \eqref{DecompRootsPND}. More precisely, we assume that this component does
not vanish asymptotically faster than a typical field component, namely that
${\fieldP{\refT}{2\!s}{}_*\neq0}$, see \eqref{fieldsnearscri}. The
vanishing coefficient ${\fieldP{\refT}{2\!s}{}_*}$ indicates that the reference
tetrad is asymptotically aligned along some PND. Indeed, considering the fact that by interchanging
${\kO}$ with ${\lO}$ the component ${\fieldP{\refT}{0}}$ goes to
${\bfieldP{\refT}{2\!s}}$, the condition ${\fieldP{\refT}{2\!s}{}_*=0}$ implies that the vector ${\lO}$
of the reference tetrad is the PND, say ${\kG_1}$. In terms of the directional parameter this means that ${\R_1=\infty}$.
In such a case we have to use a different normalization factor
to express the field components.
With help of the relation \eqref{zeroComp}, for ${\fieldP{\refT}{0}\neq0}$ we can write
\begin{equation}\label{FieldInterpCompleteZCN}
\fieldP{\intT}{2\!s} \lteq \frac{1}{\afp} \,\,\EPS_\refT^s\fieldP{\refT}{0}{}_*\,
\frac{\bigl(\nsgn\bar{\R}-\R_1^{-1}\bigr)\bigl(\nsgn\bar{\R}-\R_2^{-1}\bigr)\ldots\bigl(\nsgn\bar{\R}-\R_{2\!s}^{-1}\bigr)}
  {\bigl(1-\nsgn \R\bar{\R}\bigr)^s}\,  \period
\end{equation}
This expression describes the same directional dependence as the expression
\eqref{FieldInterpComplete}, it is only normalized using a different field
component.

The expression \eqref{FieldInterpComplete} is useful if ${\fieldP{\refT}{2\!s}{}_*\neq0}$,
the expression \eqref{FieldInterpCompleteZCN} is applicable when ${\fieldP{\refT}{0}{}_*\neq0}$.
In situations when ${\fieldP{\refT}{0}{}_*=\fieldP{\refT}{2\!s}{}_*=0}$,
so that \emph{both} the vectors ${\kO}$ and ${\lO}$ are PNDs,
another non-vanishing component ${\fieldP{\refT}{j}{}_*}$ has to be used for the normalization.
A particular example of  normalization using a different field component for the gravitational Petrov type {D} field
will be discussed in section \ref{ssc:spacelikesri}, see equation
\eqref{WTdirDAlter}.

\section{Discussion of the directional structure of radiation on $\scri$}
\label{sc:pattern}

In this section we will discuss the general expression
\eqref{FieldInterpComplete} for different values of ${\nsgn=-\sign{\Lambda}}$ in detail.
For practical purposes, we will restrict the
description to gravitational and electromagnetic fields; general spin-$s$ field will be
mentioned only for maximally degenerate field of algebraic type N.

\subsection{Radiation on null ${\,\scri}$}
\label{ssc:nullscri}

For \vague{Minkowskian}  conformal infinity we have ${\nsgn=0}$,
${\lO\!\propto\norm}$, and the field thus has \emph{no directional structure}.
In such a case the radiative parts of the gravitational and electromagnetic fields
\eqref{FieldInterpComplete} are uniquely given by
\begin{equation}\label{FieldsOnNullScri}
|\WTP{\intT}{4}| \lteq \frac{|\WTP{\refT}{4}{}_*|}{|\,\afp\,|} \comma\qquad
|\EMP{\intT}{2}| \lteq \frac{|\EMP{\refT}{2}{}_*|}{|\,\afp\,|} \commae
\end{equation}
i.e., they are \emph{the same for all null geodesics approaching a given point} ${P\in\scri}$.
For (locally) asymptotically flat spacetimes it is thus possible to  distinguish
between the radiative and non-radiative fields.
Radiation is absent at those points of null
conformal infinity where the constants $\WTP{\refT}{4}{}_*$ or
$\EMP{\refT}{2}{}_*$ vanish.
As we discussed, this occurs when \emph{the principal null direction is oriented along
the vector $\lO\!\propto\norm$}.
This can be viewed as an invariant characterization of the absence
of radiation near ${\scri}$.

In section \ref{ssc:dirstrrad} we suggested that for $\WTP{\refT}{4}{}_*=0$
we should use the alternative form of the directional pattern of radiation \eqref{FieldInterpCompleteZCN}.
However, in the case ${\nsgn=0}$ it reduces to
\begin{equation}\label{WTPInterp0LZCN}
\abs{\WTP{\intT}{4}} \lteq \frac{\abs{\WTP{\refT}{0}{}_*}}{\abs{\afp}} \;
 \abs{\R_1\R_2\R_3\R_4}^{-1}
\end{equation}
with one of the ${\R_n}$ infinite. We thus again obtain ${\WTP{\intT}{4} =0}$ in
the order ${\afp^{-1}}$.

\subsection{On the meaning of the peeling-off behaviour}
\label{ssc:peeling}

Let us give here some general comments concerning the character of the fields
near infinity which apply also to spacelike and timelike ${\scri}$.
Because for the Minkowskian infinity the leading term of the field
is independent of the direction along which the infinity
is approached, one tends to attribute the invariant meaning to the components
of the field with respect to the interpretation tetrad,
say, to the components ${\WTP{\intT}{j}}$ of the gravitational field.
The peeling-off behaviour ${\WTP{\intT}{j}\sim\afp^{j-5}}$ could  thus be
rephrased that the Weyl tensor becomes asymptotically of type {N} ---
only the component ${\WTP{\intT}{4}}$ \vague{survives} when one is approaching the infinity.
However, as pointed out in \cite{PenroseRindler:book}, such an
interpretation can be misleading. The peeling-off property is a consequence
of a delicate interplay between the decay behaviour \eqref{fieldsnearscri} of the field
and of the different asymptotic scaling of the vectors of the interpretation tetrad.
Consequently, the asymptotic type~{N} characterization of the Weyl tensor
\emph{is not invariant} --- the Weyl tensor of type~{N} should have
one quadruply degenerate PND which should coincide with the vector ${\kI}$
of the interpretation tetrad, i.e., with the vector tangent to the
geodesic along which the infinity is approached. But this vector
obviously depends on our choice and cannot thus be invariant
characterization of the Weyl tensor.

The invariant asymptotic algebraic characterization of the field
(asymptotic PNDs of the field) can be obtained by the conformal technique.
As discussed already in section \ref{ssc:PND}, PNDs do not depend on isotropic rescaling
of the field and they can thus be defined using the leading term of the
field tensor, i.e., using the field components with respect to the reference
tetrad (or any other tetrad) which is related by an \emph{isotropic} rescaling
to a tetrad well defined in the sense of the conformal manifold ${\cmfld}$.
Defining PNDs in this way, the field can be of a \emph{general} type up to the infinity.
The PNDs defined at infinity can be used to define canonical reference tetrads
as will be done in sections~\ref{ssc:spacelikesri} and \ref{ssc:timelikescri}.
For example, the ${C}$-metric spacetime is of Petrov type {D} everywhere,
including at infinity, and its double degenerate PNDs at ${\scri}$ have been used
to define the reference tetrad in \cite{KrtousPodolsky:2003}.

Because the interpretation tetrad is \emph{not} of the type described above
(the vectors ${\kI}$ and ${\lI}$ scale differently with respect to the conformal manifold),
the field components in this tetrad can exhibit apparent degeneracy
typical for type {N} fields.

As we have found, the asymptotic algebraic structure of the field allows us to give a clear
unambiguous characterization of the field near Minkowskian (null) infinity. The leading
radiative term (along any null geodesic approaching ${\scri)}$ disappears
if a PND is tangent to infinity. For spacelike and timelike infinities
the leading term depends on a direction along which ${\scri}$ is approached,
and it is absent only along some specific directions, given again by the orientation
of PNDs as described in detail in the following sections.

The invariant characterization of the absence of radiation using PNDs
raises a question of the relation between the algebraic structure
of fields (orientation of PNDs on ${\scri}$) and the structure of sources.
For example, in the case of two accelerated black holes (the $C$-metric) the two (double degenerate) PNDs
play a role of \vague{radial} directions from the holes,
cf.\ Ref.~\cite{KinnersleyWalker:1970,KrtousPodolsky:2003}.
It would be interesting to discover a similar relation between PNDs and sources
in a more general situation. We will analyze this question in another work.

\subsection{Parametrization of directions by (pseudo-)spherical angles}
\label{ssc:SpherAngles}
In order to characterize  more lucidly the directions on spacelike
or timelike $\scri$  it~is convenient to express the complex
directional parameter $\R$ in terms of \mbox{(pseudo-)}spherical parameters.

At any point ${P\in\scri}$ we have a reference null tetrad ${\kO,\,\lO,\,\mO,\,\bO}$
which is adjusted to conformal infinity. Such a tetrad is associated
with an orthonormal adjusted tetrad ${\tO,\,\qO,\,\rO,\,\sO}$,
where $\tO$ is a unit timelike vector and ${\qO,\,\rO,\,\sO}$ are
perpendicular spacelike unit vectors,
\begin{equation}\label{NullNormTetr}
\begin{aligned}
  \tO &= \textstyle{\frac1{\sqrt{2}}} (\kO+\lO)\comma&
  \qO &= \textstyle{\frac1{\sqrt{2}}} (\kO-\lO)\commae\\
  \rO &= \textstyle{\frac1{\sqrt{2}}} (\mO+\bO)\comma&
  \sO &= \textstyle{\frac{i}{\sqrt{2}}}(\mO-\bO)\commae
\end{aligned}
\end{equation}
see \eqref{NormNullTetr}. From the coplanarity and normalization
condition  \eqref{AdjustedRefer} it follows that
\begin{equation}
\begin{cases}
   \ \tO\,=\ \EPS_\refT\,\norm  & \text{\ when $\scri\,$ is spacelike} \quad  (\nsgn=-1)\commae\cr
   \ \qO  = -\EPS_\refT\,\norm  & \text{\ when $\scri\,$ is timelike}\ \quad  (\nsgn=+1)\commae\cr
\end{cases}
\label{RefNormals}\end{equation}
where $\norm$ is the normal to $\scri$, cf.~figure~\ref{fig:adjtetr}.
We can now project a null vector $\kG$, whose direction is represented
by the parameter ${\R}$ by \eqref{Rmeaning}, onto the corresponding conformal infinity.

In spacetimes with ${\Lambda>0}$, for which $\scri$ is spacelike,
we perform a normalized spatial projection to a three-dimensional space
orthogonal to $\tO$,
\begin{equation}\label{genprojq}
\qG=\frac{\kG+(\kG\spr\tO)\,\tO}{|\kG\spr\tO|}\commae
\end{equation}
where ${\kG\spr\tO=\mtrc_{ab}\kG^a\tO^b}$.
The unit spatial direction $\qG$ corresponding to $\kG$ can be expressed in
terms of standard \defterm{spherical angles} ${\THT,\,\PHI}$, with respect
to the reference tetrad,
\begin{equation}\label{THTPHIdef}
  \qG = \cos\THT\;\qO + \sin\THT\,(\cos\PHI\;\rO + \sin\PHI\;\sO)\period
\end{equation}
Substituting \eqref{Rmeaning} into \eqref{genprojq}, and comparing with
\eqref{THTPHIdef} we obtain
\begin{equation}\label{Rdef}
  R = \tan\frac{\THT}{2}\,\exp(-i\PHI)\period
\end{equation}
Therefore, $\R$ is exactly the \defterm{stereographic representation}
of the angles ${\THT,\,\PHI}$.
Additionally, for ${\nsgn=-1}$ the orientation of the null vector ${\kG}$
with respect to ${\scri}$ coincides with the orientation of ${\kO}$,
${\EPS=\EPS_\refT}$, cf.\ figure~\ref{fig:adjtetr}.

\begin{figure}
\begin{center}
\includegraphics{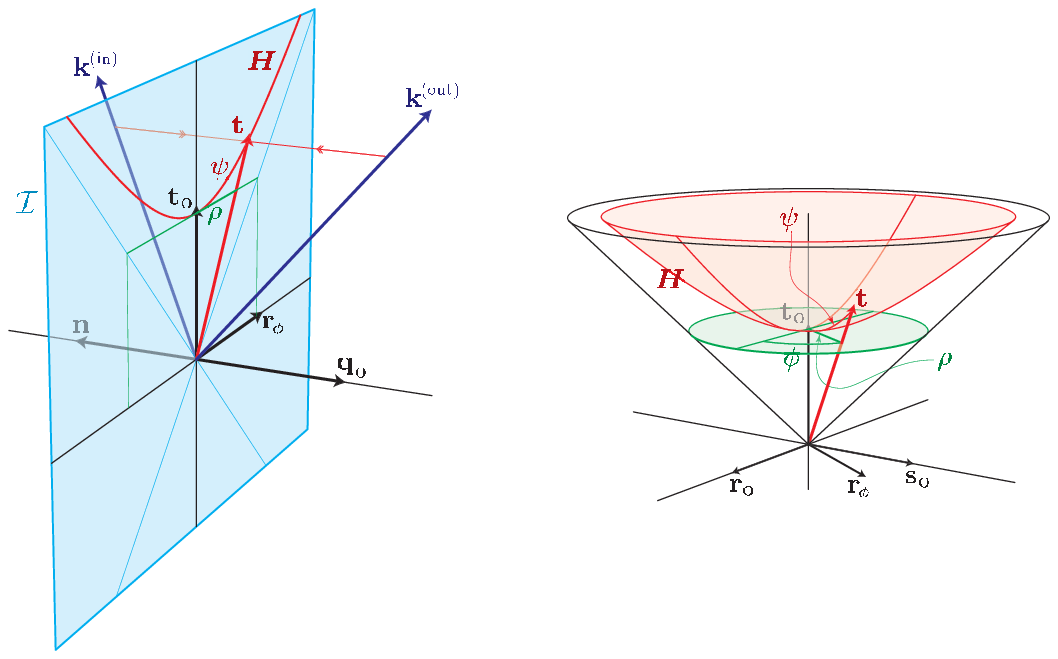}
\end{center}
\caption{\label{fig:nulldir}%
Parametrization of null directions $\kG$ near timelike infinity $\scri$.
All null directions form three families: \defterm{outgoing}  (${\EPS=+1}$,
vector $\kG^{(\mathrm{out})}$ in the figure), \defterm{ingoing} (${\EPS=-1}$,
vector~$\kG^{(\mathrm{in})}$), and directions \defterm{tangent} to~$\scri$.
The direction $\kG$ can be parameterized with respect to a reference tetrad
${\tO,\,\qO,\,\rO,\,\sO}$  by the boost~$\PSI$, angle $\PHI$ and orientation $\EPS$,
or by a complex number~$\R$, or by parameters $\RHO$, $\PHI$.
In the left diagram, the vectors ${\tO,\,\qO,\,\rG_\PHI}$, where
${\rG_\PHI=\cos\PHI\;\rO + \sin\PHI\;\sO}$, are depicted; in the right the direction
${\qO=-\EPS_\refT\norm}$ is omitted. The parameters ${\PSI,\,\PHI}$ specify the normalized
orthogonal projection $\tG$ of $\kG$ into $\scri$, cf.\ Eqs.~\eqref{genproj}, \eqref{PSIPHIparam}.
To parametrize $\kG$ uniquely, we have to specify also its orientation $\EPS$ with respect to $\scri$.
The parameter $\R$ is the Lorentzian stereographic representation
of ${\PSI,\,\PHI,\,\EPS}$, cf.\ Eq.~\eqref{RPSIPHIrel}.
Vectors~$\tG$ corresponding to all outgoing (or ingoing) null directions form
a hyperbolic surface $\boldsymbol{H}$. This can be radially mapped onto a two-dimensional
disk tangent to the hyperboloid at $\tO$,
which can be parametrized by an angle $\PHI$ and a radial coordinate ${\RHO=\tanh\PSI}$.
In the exceptional case ${\RHO=1}$, i.e. ${\PSI\to\infty}$, the vector ${\kG\propto\tG+\rG_\PHI}$ is
tangent to $\scri$. }
\end{figure}

Alternatively, in spacetimes with ${\Lambda<0}$ for which $\scri$ is timelike
the normalized projection of  $\kG$ onto $\scri$ is
\begin{equation}\label{genproj}
\tG=\frac{\kG-(\kG\spr\qO)\,\qO}{\abs{\kG\spr\qO}}\period
\end{equation}
The resulting unit timelike vector $\tG$ is tangent to the
Lorentzian (1+2) conformal infinity.
We can analogously characterize $\tG$
(and thus $\kG$) with respect to the reference tetrad as
\begin{equation}\label{PSIPHIparam}
  \tG = \cosh\PSI\;\tO + \sinh\PSI\,(\cos\PHI\;\rO + \sin\PHI\;\sO) \period
\end{equation}
The parameters ${\PSI,\PHI}$ are \defterm{pseudo-spherical} parameters,
${\PSI\in(0,\infty)}$ corresponding to a boost,
and ${\PHI\in(-\pi,+\pi)}$ being an angle.
Their geometrical meaning is visualized in figure~\ref{fig:nulldir}.
However, these parameters do not specify the null direction $\kG$ uniquely ---
there always exist \emph{one ingoing} and \emph{one outgoing} null directions with the same
parameters $\PSI$ and $\PHI$, which are distinguished by ${\EPS=\pm1}$.
Substituting equation~\eqref{Rmeaning} into \eqref{genproj},
and comparing with equation~\eqref{PSIPHIparam} we express $\PSI$ and $\PHI$
in terms of $\R$ as
\begin{equation}\label{PSIPHIRrel}
  \tanh\PSI=\frac{2\abs{\R}}{1+\abs{\R}^2}\comma\PHI=-\arg\R\period
\end{equation}
Observing that ${\sign(1-\abs{\R}^2)}$ characterizes a difference in
orientations of the vectors ${\kG}$ and ${\kO}$ with respect
to  infinity, ${\EPS=\EPS_\refT\sign(1-\abs{\R}^2)}$,
we can write down the inverse relations,
\begin{equation}\label{RPSIPHIrel}
  \R =
\begin{cases}
{\displaystyle \tanh\frac\PSI2\;\exp(-i\PHI)}&   \quad\text{for}\quad {\EPS\!=\!+\EPS_\refT}\commae
\vspace*{6pt}\\
{\displaystyle \coth\frac\PSI2\;\exp(-i\PHI)}&   \quad\text{for}\quad {\EPS\!=\!-\EPS_\refT}\period
\end{cases}
\end{equation}
We also allow an infinite value ${\R=\infty}$ which corresponds to
${\PSI=0}$, ${\EPS=-\EPS_\refT}$, i.e., ${\kG\propto(\tO-\qO)/\sqrt2}$.

Of course such a parametrization can be applied to any null direction $\kG$. In particular,
it may characterize the direction $\kI$ of a null geodesic along which the infinity is approached,
and also describe the principal null directions. The PNDs on a \vague{de~Sitter-like} $\scri$ are thus given
by the spherical angles ${\THT_n,\,\PHI_n}$ related to $\R_n$  by equation \eqref{Rdef}, whereas
on an \vague{anti--de~Sitter-like} $\scri$ by ${\PSI_n,\,\PHI_n,\,\EPS_n}$ which are given by \eqref{RPSIPHIrel}.

\subsection{Radiation on spacelike ${\scri}$}
\label{ssc:spacelikesri}
The asymptotic structure of gravitational and electromagnetic fields evaluated in
the interpretation tetrad near a \emph{de~Sitter-like conformal infinity}, ${\nsgn=-1}$
with ${\norm=\EPS_\refT\tO}$, is given by \eqref{FieldInterpComplete}
for ${s=2}$ and ${s=1}$, respectively,
\begin{align}
\WTP{\intT}{4} &\lteq \,\frac{ \,\WTP{\refT}{4}{}_*}{\afp} \,{\bigl(1+|\R|^2\bigr)^{-2}}\,
  \biggl(1-\frac{\R_1}{\R\ant}\biggr)\biggl(1-\frac{\R_2}{\R\ant}\biggr)
  \biggl(1-\frac{\R_3}{\R\ant}\biggr)\biggl(1-\frac{\R_4}{\R\ant}\biggr)\commae
  \label{WTPintT}\\
\EMP{\intT}{2} &\lteq \EPS_0\frac{\,\EMP{\refT}{2}{}_*}{\afp} \,{\bigl(1+|\R|^2\bigr)^{-1}}\,
 \biggl(1-\frac{\R_1}{\R\ant}\biggr)\biggl(1-\frac{\R_2}{\R\ant}\biggr) \commae\label{EMPintT}
\end{align}
where, using \eqref{Rdef},
\begin{equation}\label{prefactor}
  {\bigl(1+|\R|^2\bigr)^{-1}} = \,\cos^2\Bigl(\frac{\THT}{2}\Bigr)\commae
\end{equation}
and the complex number $\R\ant$ is
\begin{equation}\label{antdef}
  \R\ant = - \frac{1}{\bar{\R}} = -\cot\Bigl(\frac{\THT}{2}\Bigr)\,\exp(-i\PHI)\period
\end{equation}
It characterizes a spatial direction \defterm{opposite} to the direction given by $\R$,
i.e., the \defterm{antipodal} direction with ${\THT\ant=\pi-\THT}$ and ${\PHI\ant=\PHI+\pi}$.
The remaining freedom in the choice of the vectors ${\mI,\,\bI}$ changes just a phase
of the field components, so that only their modulus ${|\WTP{\intT}{4}|}$ or ${|\EMP{\intT}{2}|}$
has an invariant meaning.

In a general spacetime there
exist \emph{four} spatial directions at ${P\in\scri}$ along which the
radiative component of the gravitational field \eqref{WTPintT} \emph{vanishes},
namely the directions satisfying ${\R\ant = \R_n\,}$, $n=1,2,3,4$
(or \emph{two} such directions for electromagnetic field \eqref{EMPintT}).
These privileged null directions $\kG$ are given by \eqref{Rmeaning} with
${\R=(\R_n)\ant}$. Spatial parts of them are thus exactly \emph{opposite to
the projections of the principal null directions  onto~$\scri$}.

In \emph{algebraically special} spacetimes some PNDs coincide, and
expressions \eqref{WTPintT}, \eqref{EMPintT} simplify.
Moreover, it is always possible to choose the
\emph{canonical reference tetrad} aligned to the algebraic structure:
\begin{enumerate}
\item{} the vector $\qO$ is oriented along the spatial projection of the
\emph{degenerate} (multiple) PND onto $\scri$, say  ${\kG_4}$, i.e.  ${\kO =\kG_4\,}$,
\item{} the ${\qO\textdash\rO}$ plane is oriented so that it contains the spatial
projection of one of the remaining PNDs, say $\kG_1$ (for type~{N} spacetimes this choice is
arbitrary).
\end{enumerate}
Using such a canonical  reference tetrad, the degenerate PND $\kG_4$ is
parametrized  by ${\THT_4=0}$, i.e.\ ${R_4 = 0}$, see equations \eqref{THTPHIdef} and
\eqref{Rdef}. The PND $\kG_1$ has ${\PHI_1=0}$, i.e.\ ${R_1 = \tan(\THT_1/2)}$
is a real constant.

\begin{figure}
\begin{center}
\includegraphics{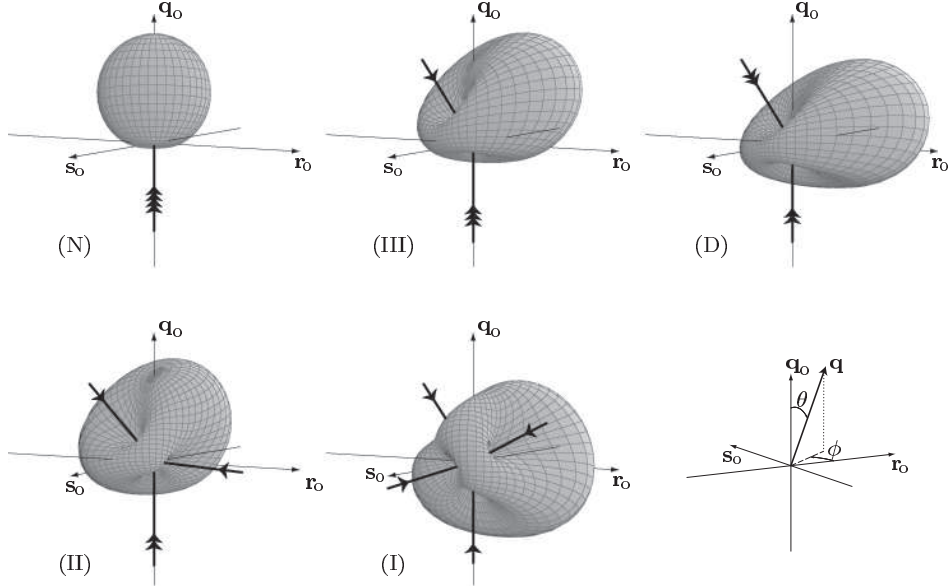}
\end{center}
\caption{\label{fig:dpr-dS}%
Specific directional structure of radiation for spacetimes of Petrov types
{N}, {III}, {D}, {II} and {I}. Directions in the diagrams are spatial
directions tangent to a spacelike $\scri$.
For each type, the radiative component $\abs{\WTP{\intT}{4}}$ along a null geodesic
is depicted in the corresponding spatial direction $\qG$
parametrized by spherical angles $\THT$, $\PHI$, see \eqref{THTPHIdef}.
[Degenerate] principal null directions (PNDs) are indicated by [multiple] bold arrows.
Thick lines represent spatial directions (opposite to PNDs)
along which the radiation vanishes.
}
\end{figure}

Consequently, for the Petrov \emph{type~{N}} spacetimes  (which have a quadruply degenerate PND),
in the canonical reference tetrad there is ${\R_1=\R_2=\R_3=\R_4=0}$, so that the
asymptotic behaviour of  gravitational field \eqref{WTPintT} becomes
\begin{equation}\label{WTdirN}
\abs{\WTP{\intT}{4}} \lteq \, \abs{\WTP{\refT}{4}{}_*}\,|\afp|^{-1}\,\cos^4\!{\textstyle\frac\THT2}\period
\end{equation}
The corresponding directional structure of radiation is illustrated
in figure~\ref{fig:dpr-dS}(N). It is axisymmetric,
with maximum value at ${\THT=0}$ along the spatial
projection of the quadruple PND onto $\scri$. Along the opposite direction,
${\THT=\pi}$, the field vanishes. Analogously, for a  spin-$s$
field of type N (with all PNDs coinciding) we obtain%
\begin{equation}\label{fielddirN}
\abs{\fieldP{\intT}{2\!s}} \lteq \, \abs{\fieldP{\refT}{2\!s}{}_*}\,|\afp|^{-1}\,\abs{\cos{\textstyle\frac\THT2}}^{2s}\period
\end{equation}

In  Petrov \emph{type~{III}} spacetimes,
${\R_1=\tan\frac{\THT_1}{2}}$, ${\R_2=\R_3=\R_4=0}$, and \eqref{WTPintT}  implies
\begin{equation}\label{WTdirIII}
\abs{\WTP{\intT}{4}} \lteq \,\abs{\WTP{\refT}{4}{}_*}\,|\afp|^{-1}\,\cos^4\!{\textstyle\frac\THT2}\,
     \abs{1+\tan{\textstyle\frac{\THT_1}2}\,\tan{\textstyle\frac\THT2}\,e^{i\PHI}}\period
\end{equation}
This directional pattern is shown in figure~\ref{fig:dpr-dS}(III).
The field vanishes along ${\THT=\pi}$ and along
${\THT=\pi-\THT_1}$, ${\PHI=\pi}$ which are spatial directions opposite to the PNDs.

The \emph{type~{D}} spacetimes admit two double degenerate PNDs,
${\R_1=\R_2=\tan\frac{\THT_1}{2}}$
and ${\R_3=\R_4=0}$.
The gravitational  field near spacelike $\scri$ thus takes the form%
\begin{equation}\label{WTdirD}
\abs{\WTP{\intT}{4}} \lteq \,\abs{\WTP{\refT}{4}{}_*}\,|\afp|^{-1}\,\cos^4\!{\textstyle\frac\THT2}\,
    \abs{1+\tan{\textstyle\frac{\THT_1}2}\,\tan{\textstyle\frac\THT2}\,e^{i\PHI}}^2\commae
\end{equation}
with two planes of symmetry, see figure~\ref{fig:dpr-dS}(D).
This directional dependence agrees with that for the
$C$-metric spacetime with ${\Lambda>0}$ derived recently in \cite{KrtousPodolsky:2003}.

For Petrov \emph{type~{II}} spacetimes, only two PNDs coincide so that
${\R_1=\tan\frac{\THT_1}{2}}$, ${\R_2=\tan\frac{\THT_2}{2}\exp(-i\PHI_2)}$,
${\R_3=\R_4=0}$. Asymptotic directional structure of the field,
\begin{equation}\label{WTdirII}
\abs{\WTP{\intT}{4}} \lteq \,\abs{\WTP{\refT}{4}{}_*}\,|\afp|^{-1}\,\cos^4\!{\textstyle\frac\THT2}\,
    \bigl\lvert{1+\tan{\textstyle\frac{\THT_1}2}\,\tan{\textstyle\frac\THT2}\,e^{i\PHI}}\bigr\rvert
    \bigl\lvert{1+\tan{\textstyle\frac{\THT_2}2}\,\tan{\textstyle\frac\THT2}\,e^{i(\PHI-\PHI_2)}}\bigr\rvert\commae
\end{equation}
is drawn in figure~\ref{fig:dpr-dS}(II).

Finally, in case of algebraically general \emph{type~{I}}
spacetimes one needs five real parameters
$\THT_1$, $\THT_2$, $\PHI_2$, $\THT_3$, $\PHI_3$
to characterize the directional dependence
\begin{equation}\label{WTdirI}
\begin{aligned}
\abs{\WTP{\intT}{4}} \lteq \,&\abs{\WTP{\refT}{4}{}_*}\,|\afp|^{-1}\,\cos^4\!{\textstyle\frac\THT2}\,
    \bigl\lvert{1+\tan{\textstyle\frac{\THT_1}2}\,\tan{\textstyle\frac\THT2}\,e^{i\PHI}}\bigr\rvert   \\
& \qquad\quad\times\bigl\lvert{1+\tan{\textstyle\frac{\THT_2}2}\,\tan{\textstyle\frac\THT2}\,e^{i(\PHI-\PHI_2)}}\bigr\rvert
        \bigl\lvert{1+\tan{\textstyle\frac{\THT_3}2}\,\tan{\textstyle\frac\THT2}\,e^{i(\PHI-\PHI_3)}}\bigr\rvert\commae
\end{aligned}
\end{equation}
figure~\ref{fig:dpr-dS}(I), of the gravitational field with respect to the canonical reference tetrad.

Of course, for any \emph{conformally flat} spacetime the radiation vanishes entirely because
${\WTP{\intT}{j}=0}$ for all $j$. This is the case of,
for example, the Friedman-Robertson-Walker solutions which admit ${\scri}$.

There exist alternative choices of the reference tetrad, e.g., those which respect the
symmetry of the radiation pattern. For spacetimes  of \emph{type~{D}} the directional structure
indicated in figure~\ref{fig:dpr-dS}(D) admits two planes of symmetry. It is thus natural to choose the
tetrad ${\qG_\refT', \rG_\refT', \sG_\refT'}$ adapted to them: we require that one (double degenerate) PND
has inclination $\THTS$ with respect to ${\qG_\refT'}$, the
second PND has the \emph{same} inclination with respect to ${-\qG_\refT'}$
(i.e.\ ${\THTS=(\pi-\THT_1)/2}$), and that the
vector ${\sG_\refT'}$ is perpendicular to the plane spanned by these PNDs
(see \cite{KrtousPodolsky:TDF} for more details).
With respect to this reference tetrad the PNDs are parameterized by the
coefficients ${R_1=R_2=\tan{\textstyle\frac\THTS2}}$ and
${R_3=R_4=\cot{\textstyle\frac\THTS2}}$.
Moreover, for type D there exists a natural normalization of the field which is different from that discussed above.
One can evaluate the components ${\WTP{\spcT}{j}}$ in the \defterm{algebraically special null tetrad} with
${\kS}$ and ${\lS}$ given by the degenerate PNDs --- it follows from the definition of PNDs that
only the component ${\WTP{\spcT}{2}}$ would be non-vanishing. This component is independent of a choice of the
tetrad vectors orthogonal to PNDs, and of the scaling of PNDs
(assuming ${\kS\spr\lS=-1}$), cf.\ \eqref{boostrotationField} with ${s=2}$.
We may thus use ${\WTP{\spcT}{2}}$ to normalize the directional
structure of radiation.
Using the relation
\begin{equation}\label{TypeDNorm}
\WTP{\refT\prime}{4}=\textstyle{\frac32}\tan^2\THTS\;\WTP{\spcT}{2}\commae
\end{equation}
see \cite{KrtousPodolsky:TDF}, the radiation pattern \eqref{WTPintT}
parametrized by angles ${\THT'}$, ${\PHI'}$ with respect to the
reference tetrad ${\tG_\refT', \qG_\refT', \rG_\refT', \sG_\refT'}$ reads
\begin{equation}\label{WTdirDAlter}
\abs{\WTP{\intT}{4}} \lteq \frac{1}{\abs{\afp}}
\,\frac32\frac{\abs{\WTP{\spcT}{2}{}_*}}{\cos^2\THTS}\,
  \bigl\lvert \sin\THT'+\sin\THTS\cos\PHI'-i \sin\THTS\cos\THT'\sin\PHI'\bigr\rvert^2\period
\end{equation}
This coincides with  the expression for the asymptotic directional structure of radiation in
the $C$-metric spacetime with ${\Lambda>0}$, as previously presented in \cite{KrtousPodolsky:2003}.

For a completely general choice of the reference tetrad near a de~Sitter-like conformal infinity,
the dominant radiative term \eqref{WTPintT} of any gravitational field can asymptotically be written in terms of spherical angles ${\THT}$, ${\PHI}$ as
\begin{equation}\label{WTdirIgen}
\abs{\WTP{\intT}{4}} \lteq \,\frac{\abs{\WTP{\refT}{4}{}_*}}{|\afp|}\,\,\cos^4{\textstyle\frac\THT2}\,
  \mspace{-13mu}\prod_{n=1,2,3,4}\mspace{-10mu}  \bigl\lvert{1+\tan{\textstyle\frac{\THT_n}2}\,\tan{\textstyle\frac\THT2}\,e^{i(\PHI-\PHI_n)}}\bigr\rvert\commae
\end{equation}
where ${\THT_n, \PHI_n}$ identify the principal null directions  $\kG_n$ with respect to the
reference tetrad. In a similar way, when ${\WTP{\refT}{4}{}_*=0}$,
${\WTP{\refT}{0}{}_*\neq0}$ we obtain from \eqref{FieldInterpCompleteZCN}
\begin{equation}\label{WTdirIgenZCN}
\begin{split}
\abs{\WTP{\intT}{4}}
&\lteq \,\frac{\abs{\WTP{\refT}{0}{}_*}}{\abs{\afp}}\,\,\bigabs{1+\abs{\R\ant}^2}^{-2}
\abs{1-\frac{\R_1{}\ant}{\R}}\abs{1-\frac{\R_2{}\ant}{\R}}\abs{1-\frac{\R_3{}\ant}{\R}}\abs{1-\frac{\R_4{}\ant}{\R}}\\
&=\,\frac{\abs{\WTP{\refT}{0}{}_*}}{\abs{\afp}}\,\,\sin^4{\textstyle\frac\THT2}\,
  \mspace{-13mu}\prod_{n=1,2,3,4}\mspace{-10mu}
  \bigl\lvert{1+\cot{\textstyle\frac{\THT_n}2}\,\cot{\textstyle\frac\THT2}\,e^{i(\PHI-\PHI_n)}}\bigr\rvert\period
\end{split}
\end{equation}

An analogous discussion also applies to electromagnetic field \eqref{EMPintT}.
Moreover, it turns out that the square of ${\EMP{\intT}{2}}$ is the magnitude
of the Poynting vector with respect to the interpretation tetrad,
${\abs{\EMS_\intT}\lteq\frac1{4\pi}\abs{\EMP{\intT}{2}}^2}$. If the two PNDs
of the electromagnetic field coincide (${R_1=R_2=0}$)
the directional dependence of the Poynting vector at $\scri$ with respect to
the canonical reference tetrad is the same as in Eq.~\eqref{WTdirN}, figure~\ref{fig:dpr-dS}(N).
If they differ  (${R_1=\tan\frac{\THT_1}{2}}$, ${R_2=0}$),
the  asymptotic directional structure of ${\abs{\EMS_\intT}}$ is given by Eq.~\eqref{WTdirD},
illustrated  in figure~\ref{fig:dpr-dS}(D). The latter result was first  obtained for
the test field of uniformly accelerated charges in de~Sitter spacetime \cite{BicakKrtous:2002}
and then recovered in the context of the charged $C$-metric spacetime \cite{KrtousPodolsky:2003}.

The above discussion and explicit forms of the radiative directional
patterns apply both to future conformal infinity $\scri^+$ and
past $\scri^-$. In particular, it means that not only \emph{outgoing}
radiation does not vanish in a generic direction, but also that the \emph{ingoing}
field has a radiative (${\sim\afp^{-1}}$) term along a generic null geodesic coming
from the past infinity. This result can be related to Penrose's discussion of
nature of incoming field near a spacelike infinity \cite{Penrose:1964,Penrose:1967}
which has been studied in more detail in \cite{BicakKrtous:2001} and identified
as the insufficiency of purely retarded fields.

\subsection{Radiation on timelike ${\scri}$}
\label{ssc:timelikescri}

Now we shall explicitly analyze the dependence of  radiation
on the direction of a null geodesic near the \vague{anti--de~Sitter-like}, i.e.\
timelike, conformal infinity \cite{KrtousPodolsky:2004a}. With respect to a suitable reference
tetrad ${\tO,\,\qO,\,\rO,\,\sO}$ these directions are parametrized  by
the complex parameter $\R$, or its \vague{Lorentzian angles}
$\PSI$, $\PHI$ and the orientation $\EPS$, related to $\R$ by
pseudo-stereographic representation, see \eqref{RPSIPHIrel} and
figure~\ref{fig:nulldir}. The directional structure of radiation
is given by expression \eqref{FieldInterpComplete} for $\nsgn=+1$,
\begin{align}
\begin{split}
\WTP{\intT}{4} &\lteq \,\frac{  \WTP{\refT}{4}{}_*}{\afp} \,{\bigl(1-|\R|^2\bigr)^{-2}}\,
  \biggl(1-\frac{\R_1}{\R\mir}\biggr)\biggl(1-\frac{\R_2}{\R\mir}\biggr)
  \biggl(1-\frac{\R_3}{\R\mir}\biggr)\biggl(1-\frac{\R_4}{\R\mir}\biggr)\commae
\end{split}\label{WTPintTa}\\
\EMP{\intT}{2} &\lteq \EPS_0\,\frac{\EMP{\refT}{2}{}_*}{\afp} \,{\bigl(1-|\R|^2\bigr)^{-1}}\,
 \biggl(1-\frac{\R_1}{\R\mir}\biggr)\biggl(1-\frac{\R_2}{\R\mir}\biggr) \period\label{EMPintTa}
\end{align}
Here, the complex number $\R\mir$ is
\begin{equation}\label{mirdef}
  \R\mir = {\bar R}^{-1} = \coth^{\EPS\EPS_\refT}\Bigl(\frac{\PSI}{2}\Bigr)\,\exp(-i\PHI)\commae
\end{equation}
see \eqref{RPSIPHIrel}. It characterizes a direction obtained from the direction $\R$
by a \emph{reflection with respect to $\scri$}, i.e., the \defterm{mirrored}
direction with ${\PSI\mir=\PSI}$, ${\PHI\mir=\PHI}$ but opposite orientation ${\EPS\mir=-\EPS}$.
Near an anti--de~Sitter-like conformal infinity, a generic gravitational field thus takes the
asymptotic form
\begin{equation}\label{WTdirIgena}
\abs{\WTP{\intT}{4}} \lteq \,\frac{\abs{\WTP{\refT}{4}{}_*}}{|\afp|}\,\,\Bigl(\frac{\cosh\PSI+\EPS\EPS_\refT}{2}\Bigr)^2
  \mspace{-13mu}\prod_{n=1,2,3,4}\mspace{-10mu}  \bigl\lvert{1-\tanh^{\EPS_n\EPS_\refT}\!\bigl({\textstyle\frac{\PSI_n}2}\bigl)
   \,\tanh^{\EPS\EPS_\refT}\!\bigl({\textstyle\frac\PSI2}\bigl)\,e^{i(\PHI-\PHI_n)}}\bigr\rvert\commae
\end{equation}
where ${\PSI_n, \PHI_n, \EPS_n}$ identify the principal null directions  $\kG_n$, including their orientation
with respect to ${\scri}$.

The expression \eqref{WTPintTa} has been derived assuming ${\WTP{\refT}{4}\neq0}$, i.e.,
${\R_n\neq\infty}$. However, to describe the PND oriented along $\lO$ it is necessary
to use a different component $\WTP{\refT}{j}$ as a normalization factor.
With ${\WTP{\refT}{0}\!=\!\WTP{\refT}{4}\R_1\R_2\R_3\R_4}$ we obtain
\begin{equation}
\begin{split}
\abs{\WTP{\intT}{4}}
&\lteq\,\frac{\abs{\WTP{\refT}{0}{}_*}}{\abs{\afp}}\,\bigabs{1-\abs{\R\mir}^2}^{-2}
\abs{1-\frac{\R_1{}\mir}{\R}}\abs{1-\frac{\R_2{}\mir}{\R}}\abs{1-\frac{\R_3{}\mir}{\R}}\abs{1-\frac{\R_4{}\mir}{\R}}\\
&= \,\frac{\abs{\WTP{\refT}{0}{}_*}}{\abs{\afp}}\,\,\Bigl(\frac{\cosh\PSI-\EPS\EPS_\refT}{2}\Bigr)^2
  \mspace{-13mu}\prod_{n=1,2,3,4}\mspace{-10mu}  \bigl\lvert{1-\coth^{\EPS_n\EPS_\refT}\!\bigl({\textstyle\frac{\PSI_n}2}\bigl)
   \,\coth^{\EPS\EPS_\refT}\!\bigl({\textstyle\frac\PSI2}\bigl)\,e^{i(\PHI-\PHI_n)}}\bigr\rvert
\period
\end{split}\label{WTPintTmir}
\end{equation}
Interestingly, the radiation pattern has thus the same form if
all PNDs are reflected, ${\R_n\to(\R_n)\mir}$, and ingoing and
outgoing directions switched, ${\R\to\R\mir}$.

Both expressions \eqref{WTPintTa} and \eqref{WTPintTmir} characterize the asymptotic
behaviour of the fields near anti--de~Sitter-like
infinity.  First, we observe
from \eqref{WTPintTa} that
the radiation \vague{blows up} for directions with ${\abs{\R}\!=\!1}$
(i.e., ${\PSI\to\infty}$). These are null directions \emph{tangent} to
$\scri$, and thus they do not represent a direction
of any geodesic approaching $\scri$ from
the \vague{interior} of  spacetime. The reason for this divergent behaviour
is \vague{kinematic}: when we required the \vague{comparable} approach of geodesics
to infinity (see discussion nearby \eqref{kIfixing})
we had fixed the component of $\kI$ normal to $\scri$, equation \eqref{kAscaling}.
Clearly, such a condition implies an \vague{infinite} rescaling if
$\kI$ is tangent to $\scri$ which results in the divergence of $\abs{\WTP{\intT}{4}}$.

\begin{figure}
\begin{center}
\includegraphics{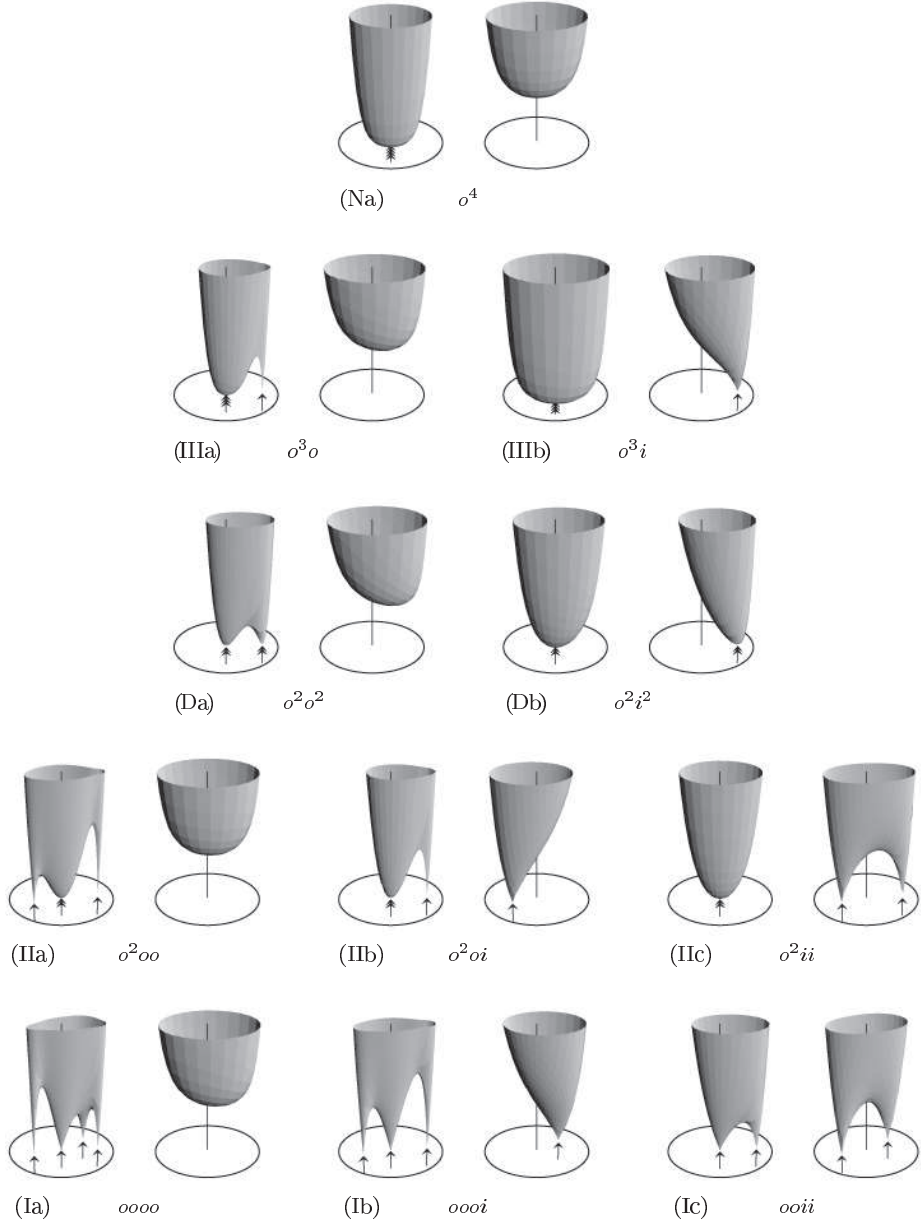}
\end{center}
\caption{\label{fig:dpr-AdS}%
Directional structure of radiation near a timelike $\scri$. All 11 qualitatively
different shapes of the pattern when PNDs are not tangent to $\scri$ are shown
(remaining 9 are related by a simple reflection with respect to $\scri$).
Each diagram consists of patterns for ingoing (left) and outgoing geodesics
(right). $\abs{\WTP{\intT}{4}}$ is drawn on the vertical axis, directions of
geodesics are represented on the horizontal disc by coordinates ${\RHO,\,\PHI}$
introduced in figure~\ref{fig:nulldir}. \emph{Reflected} [degenerated] PNDs are
indicated by [multiple] arrows under the discs. For PNDs that are not tangent
to $\scri$ these are directions of vanishing radiation. The Petrov types
(N,~III, D, II, I) corresponding to the degeneracy of PNDs are indicated by
labels of diagrams, number of ingoing and outgoing PNDs is also displayed using notation of table \ref{tab}. }
\end{figure}

\begin{figure}
\begin{center}
\includegraphics{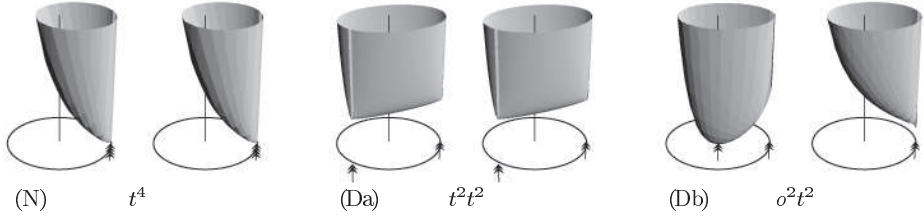}
\end{center}
\caption{\label{fig:dpr-AdS-T}%
Examples of directional structure of radiation near a timelike $\scri$ when PNDs are tangent to ${\scri}$.
Only the patterns for type N and D are shown. The notation and meaning of the
diagrams are the same as in figure \ref{fig:dpr-AdS}. }
\end{figure}

The divergence at ${\abs{\R}\!=\!1}$ splits the radiation pattern
into two components---the pattern for \emph{outgoing} geodesics
(${\EPS=+1}$) and that for \emph{ingoing} geodesics (${\EPS=-1}$).
These two different patterns are separately depicted
in figures~\ref{fig:dpr-AdS} and \ref{fig:dpr-AdS-T}.

From Eq.~\eqref{WTPintTa} it is obvious that there are, in \mbox{general},
\emph{four} directions along which the radiation \emph{vanishes},
namely PNDs \emph{reflected with respect to}
$\scri$, given by ${\R\!=\!(\R_n)\mir}$. Outgoing PNDs give rise to zeros in
the radiation pattern for ingoing null geodesics, and vice versa.
A qualitative shape of the radiation pattern thus depends on
\begin{enumerate}
\item{} \emph{degeneracy} of the PNDs (Petrov type of the spacetime),
\item{} \emph{orientation} of these PNDs with respect to~$\scri$
   (the number of outgoing/tangent/in\-going principal null directions).
\end{enumerate}
Depending on these factors there are \emph{51 qualitatively different shapes of the radiation
patterns} (3 for Petrov type~N spacetimes, 9 for type~III, 6 for~D, 18 for~II, and 15 for type~I
spacetimes); 21 pairs of them are related by the duality
of Eqs.~\eqref{WTPintTa} and \eqref{WTPintTmir}. All the different
possibilities are summarized in table~\ref{tab}.
The corresponding directional patterns with PNDs not tangent to ${\scri}$
are shown in figure~\ref{fig:dpr-AdS}, some examples of
those with PNDs tangent to ${\scri}$ can be found in figure~\ref{fig:dpr-AdS-T}.

As we have said before, the reference tetrad can be chosen to capture
the geometry of the spacetime. To simplify the radiation pattern we can
also adapt it to the algebraic structure, i.e., to correlate the tetrad with PNDs,
as we did thoroughly for spacelike ${\scri}$ in the previous section.
In the case of timelike conformal infinity, however, the choice of canonical reference
tetrads adjusted to PNDs is not very transparent --- it splits to a lengthy discussion
of separate cases depending on orientation of the PNDs with respect to ${\scri}$.
We do not include such a discussion here. We will only mention the simplest
case of type N fields, and investigate in some more detail the
cases of PNDs tangent to ${\scri}$, the presence of which is specific
for spacetimes with the timelike infinity.

\begin{table}
 \[
  \begin{array}{|c||c|c|} \hline
   \text{type} &  \text{PND degeneracy} &  \text{different possible orientations of PNDs}  \\ \hline\hline
           &   & \atbox{o^4} \\
\text{N}   & 4 & \atbox{t^4} \\
           &   & \atbox{i^4} \\
  \hline
           &       & \atbox{o^3o}\quad \atbox{o^3t}\quad \atbox{o^3i} \quad \atbox{t^3o} \\
\text{III} & {3+1} & \atbox{t^3t} \\
           &       & \atbox{i^3i}\quad \atbox{i^3t}\quad \atbox{i^3o} \quad \atbox{t^3i}  \\
  \hline
           &       & \atbox{o^2o^2} \quad \atbox{o^2t^2} \\
\text{D}   & {2+2} & \atbox{o^2i^2} \quad \atbox{t^2t^2} \\
           &       & \atbox{i^2i^2} \quad \atbox{i^2t^2} \\
  \hline
           &         & \atbox{o^2oo} \quad \atbox{o^2ot} \quad \atbox{o^2oi} \quad \atbox{o^2ii} \quad \atbox{o^2it} \quad \atbox{o^2tt} \quad \atbox{t^2oo} \quad \atbox{t^2ot} \\
\text{II}  & {2+1+1} & \atbox{t^2oi} \quad \atbox{t^2tt} \\
           &         & \atbox{i^2ii} \quad \atbox{i^2it} \quad \atbox{i^2io} \quad \atbox{i^2oo} \quad \atbox{i^2ot} \quad \atbox{i^2tt} \quad \atbox{t^2ii} \quad \atbox{t^2it} \\
  \hline
           &           & \atbox{oooo}\quad \atbox{ooot}\quad \atbox{oooi}  \quad \atbox{ooit} \quad \atbox{oott} \quad \atbox{ottt} \\
\text{I}   & {1+1+1+1} & \atbox{ooii} \quad \atbox{oitt} \quad \atbox{tttt} \\
           &           & \atbox{iiii}\quad \atbox{iiit}\quad \atbox{iiio}  \quad \atbox{iiot} \quad \atbox{iitt} \quad \atbox{ittt} \\
  \hline
 \end{array}
 \]
 \caption{All 51 qualitatively different directional structures of gravitational radiation near
 a timelike conformal infinity. For various algebraic Petrov types, given by the degeneracy of
 principal null directions, the specific structure is determined by the orientation of these PNDs with respect to
 $\scri$. We denote outgoing, tangent, and ingoing PNDs by the symbols $o$, $t$, and $i$, respectively,
 and their degeneracy by the corresponding power.
 The possibilities for each Petrov type which are presented in the third line are obtained from those
 in the first line by the duality  between outgoing and ingoing  directions, i.e. by interchanging $o$ with $i$.
 }
\label{tab}
\end{table}

For type N fields with the quadruply degenerate PND, which is \emph{not} tangent to ${\scri}$,
we can align the vector ${\kO}$ along this algebraically special direction, i.e.,
${\kO=\kG_1(=\kG_2=\kG_3=\kG_4)}$.
The vector ${\lO}$ is fixed by the adjustment condition \eqref{AdjustedRefer}.
(The spatial vectors ${\mO,\,\bO}$ cannot be fixed canonically
by the algebraic structure --- they have to be specified by other means.)
The PNDs are then given by ${\R_n=0}$,
i.e., ${\PSI_n=0}$ with orientations ${\EPS_n=\EPS_\refT}$, ${n=1,\,2,\,3,\,4}$.
The directional dependence of radiation \eqref{WTdirIgena} thus reduces to
\begin{equation}\label{dprN}
\abs{\WTP{\intT}{4}}\lteq\frac{\abs{\WTP{\refT}{4}{}_*}}{\abs{\afp}}\,\,\Bigl(\frac{\cosh\PSI+\EPS\EPS_\refT}{2}\Bigr)^2\commae
\end{equation}
illustrated in figure~\ref{fig:dpr-AdS}(N).
Similarly, the radiative component of a general spin-${s}$ field of type N would be
\begin{equation}\label{fieldsdprN}
\abs{\fieldP{\intT}{2\!s}}\lteq\frac{\abs{\fieldP{\refT}{2\!s}{}_*}}{\abs{\afp}}\,\,\Bigl(\frac{\cosh\PSI+\EPS\EPS_\refT}{2}\Bigr)^s
\period
\end{equation}

It is possible to introduce naturally the reference tetrads adjusted to the algebraic
structure for Petrov type D gravitational fields or, in general, for fields with two
equivalent special algebraic directions as, e.g., for a generic electromagnetic field.
Such a tetrad is analogous to that introduced above \eqref{WTdirDAlter} near
a spacelike ${\scri}$. A detailed discussion of these tetrads and of the normalization
of the field can be found in \cite{KrtousPodolsky:TDF} (cf.\ also \eqref{dprDtang} below).

We now turn to a special situation specific for the fields near a timelike infinity ${\scri}$.
Up to now we have discussed principal null directions which are either \emph{incoming} or
\emph{outgoing} from the spacetime. However, PNDs can also be \emph{tangent} to ${\scri}$,
and in the following we will discuss the consequences of such special orientation
of PNDs for the radiation pattern.
We do not expect PNDs to be tangent to $\scri$ at generic points.
However, they can be tangent on some lower-dimensional
subspace such as the intersection of $\scri$ with Killing
horizons --- cf.\ the anti--de~Sitter $C$-metric \cite{PodolskyOrtaggioKrtous:2003}.
These subspaces can be important, e.g., as in
the context of the Randall-Sundrum model:
a brane constructed from the $C$-metric reaches  infinity
with PNDs tangent both to it and to $\scri$ \cite{Emparanetal:2000b}.

In the case when all PNDs \emph{are tangent} to the conformal infinity, ${\R_n=\exp(-i\PHI_n)}$,
the directional pattern \eqref{FieldInterpComplete} for a general spin-$s$ field reduces  to
\begin{equation}\label{gendpralltang}
\abs{\fieldP{\intT}{2\!s}}\lteq \abs{\fieldP{\refT}{2\!s}{}_*}\,\abs{\afp}^{-1}
  \prod_{n=1}^{2s}
  \bigl(\cosh\PSI-\sinh\PSI\cos(\PHI\!-\!\PHI_n)\bigr)^{1/2}\!\period
\end{equation}
The field has, in general, no directions of vanishing radiation. It can only vanish along
unphysical directions ${\R=\R_n}$ (unphysical because they are tangent to ${\scri}$),
provided the PND ${\kG_n}$ is at least triple degenerate.

For type N fields, when all PNDs are the same, we can chose the reference terad in such a way that
${\R_n=1}$, i.e., ${\PHI_n=0}$, and we obtain
\begin{equation}\label{gendprNtang}
\abs{\fieldP{\intT}{2\!s}}\lteq \abs{\fieldP{\refT}{2\!s}{}_*}\,\abs{\afp}^{-1}
  \bigl(\cosh\PSI-\sinh\PSI\cos\PHI\bigr)^s\!\period
\end{equation}
In particular, for gravitational field
\begin{equation}\label{dprNtang}
\abs{\WTP{\intT}{4}}
  \lteq \abs{\WTP{\refT}{4}{}_*}\,\abs{\afp}^{-1}
  (\cosh\PSI-\sinh\PSI\cos\PHI)^2\commae
\end{equation}
see figure~\ref{fig:dpr-AdS-T}(N).

For a gravitational field of Petrov type D with \emph{both} double degenerate PNDs tangent to ${\scri}$
(figure~\ref{fig:dpr-AdS-T}(Da)), we can choose the reference tetrad such that
${\R_1=\R_2=1}$ and ${\R_3=\R_4=-1}$. The radiation pattern then becomes
\begin{equation}\label{dprDtang}
\abs{\WTP{\intT}{4}}
  \lteq \textstyle{\frac32}\abs{\WTP{\spcT}{2}{}_*}\,\abs{\afp}^{-1}\;
  (1+\sinh^2\PSI\sin^2\PHI)\commae
\end{equation}
where for normalization we have used the only non-vanishing field component ${\WTP{\spcT}{2}}$ in the
algebraically special tetrad aligned along both PNDs: this is related to the reference tetrad
field component by ${\WTP{\refT}{4}=\frac32\WTP{\spcT}{2}}$, see \cite{KrtousPodolsky:TDF}.
As we have said, there is no direction (even an unphysical one) of vanishing radiation in this case.
However, directionally dependent limits ${\R\to\R_1}$ and ${\R\to\R_4}$, in general, do not diverge
(cf.\ figure~\ref{fig:dpr-AdS-T}(Da)).
Finally, for a gravitational field of type D with \emph{only one} PND tangent to~${\scri}$,
figure~\ref{fig:dpr-AdS-T}(Db), we can choose the reference tetrad so that
${\R_1=\R_2=1}$, ${\R_3=\R_4=0}$,
\begin{equation}\label{dprDonetang}
\abs{\WTP{\intT}{4}}
  \lteq \abs{\WTP{\refT}{4}{}_*}\,\abs{\afp}^{-1}
  \frac{\cosh\PSI+\EPS\EPS_\refT}{2}\,\bigl(\cosh\PSI-\sinh\PSI\cos\PHI\bigr)\period
\end{equation}

To summarize, when $\scri$ is not null the radiation fields
depend on the direction along which the conformal infinity is approached.
Analogously to the ${\Lambda>0}$ case \cite{KrtousPodolskyBicak:2003}
the radiation pattern for ${\Lambda<0}$ has a universal
character determined by the \emph{algebraic type} of the fields \cite{KrtousPodolsky:2004a}.
However, new features occur when ${\Lambda<0}$:
both \emph{outgoing} and \emph{ingoing} patterns have to be studied,
their shapes depend also on the \emph{orientation of the PNDs}
with respect to $\scri$, and an interesting possibility of PNDs \emph{tangent to $\scri$} appears.
Radiation vanishes only along directions which are reflections of
PNDs with respect to $\scri$.
In a \emph{generic} direction it is \emph{non-vanishing}.
The absence of ${\afp^{-1}}$ term thus cannot be used to
distinguish  nonradiative sources:
near an anti--de~Sitter-like infinity the radiative component
reflects not only properties of  sources but also
their  relation to the observer.

\section{Conclusions}
\label{sc:conclude}
The investigation of the asymptotic structure of general fields in
spacetimes with a non-vanishing cosmological constant $\Lambda$
is motivated, among other, by the fact that these spacetimes have been
commonly used in various branches of theoretical research, e.g. in
inflationary models, brane cosmologies, supergravity or string theories.
Perhaps most importantly, the possible presence of a positive $\Lambda$ is also indicated
by recent observations.

An understanding of the nature of radiation in spacetimes with a non-vanishing ${\Lambda}$
is not so developed as that in spacetimes with ${\Lambda=0}$. Standard techniques used
for asymptotically flat spacetimes (such as the Bondi-Sachs approach) cannot be
applied, and generalizations of other methods lead to results which are \vague{less unique}.
In particular, we have documented that for ${\Lambda\ne0}$ the field components with
respect to a parallelly transported interpretation tetrad depend on a null direction
along which infinity is approached --- the feature which is absent in the
${\Lambda=0}$ case. In Penrose's words (cf.\ discussion after equation (9.7.38)
in \cite{PenroseRindler:book}): ``on varying geodesic through
${P}$, the different components ${\WTP{\intT}{j}}$ get mingled with each other.''
We derived this directional structure of radiation  explicitly
and we demonstrated that it is determined by the algebraic structure of the field.
The asymptotic behaviour near~$\scri$ of the dominant component of any zero-rest-mass field of spin~$s$
is given by the formula \eqref{FieldInterpComplete},
\begin{equation}\label{FieldInterpCompleteSum}
\fieldP{\intT}{2\!s} \propto \,{\afp}^{-1} \,\,
  \bigl(1-\nsgn \R\bar{\R}\bigr)^{-s}
  \prod_{n=1}^{2\!s}  {\bigl(1-\nsgn\R_n\bar{\R}\bigr)}  \commae
\end{equation}
where $\afp$ is affine parameter. The coefficient ${\nsgn=-1,0,\text{ or }+1}$ denotes
the spacelike, null, or timelike character of the conformal infinity;
in (electro)vacuum spacetimes ${\nsgn=-\sign{\Lambda}}$.
The complex parameter $\R$ represents the direction of the outgoing/ingoing null geodesic along
which a given point ${P\in\scri}$  is approached as ${\afp\to\pm\infty}$.
The complex constants $\R_n$ characterize the principal null directions,
i.e. the algebraic structure  of the field at $P$. Obviously, for $\scri$ of
a \vague{Minkowskian} type   (${\nsgn=0}$) the directional dependence completely
vanishes. The specific dependence of $\fieldP{\intT}{2\!s}$ on the direction $\R$
of the geodesic occurs if $\nsgn\not=0$, i.e., near \vague{(anti--)de~Sitter-like}
conformal infinity. Interestingly, in all spacetimes which are not conformally flat
there are \emph{at most} $2s$ directions along
which the radiative part of the field \eqref{FieldInterpCompleteSum} vanishes. These are
directions antipodal to the principal null directions in the case of a spacelike $\scri$, and
mirror reflections of the PNDs with respect to $\scri$ when its character is timelike.
Along all other directions the radiation does \emph{not} vanish, even if the field
corresponds to a \vague{static} source.

Our results  supplement and refine the peeling-off behaviour of zero-rest-mass
fields.  The \vague{peeling} is a well-known property of the fields near
conformal infinity, and therefore we will emphasize again its relation
to the above derived asymptotic directional structure of radiation.
For example, in classical works \cite{Sachs:1962,Pirani:1965} one can find its very suggestive
formulation: the curvature tensor expanded along null geodesics takes the form
\begin{equation}\label{wrongpeeling}
\WTP{}{} = \mathrm{N}\, \afp^{-1} + \mathrm{III}\, \afp^{-2} +
\mathrm{II}\, \afp^{-3} + \mathrm{I}\, \afp^{-4}+ \dots\commae
\end{equation}
(see page 365 in \cite{Pirani:1965} or equation (5.6) in \cite{Sachs:1962})
where the terms N, III, II, and I are algebraically
special with quadruple, triple, double, and non-degenerate PNDs, respectively.
On this basis it is commonly stated that the radiative component (${\sim\afp^{-1}}$) becomes
asymptotically of Petrov type N with one quadruply degenerate PND.
Our discussion above, however, demonstrates that such an interpretation is misleading or, at
least, not precise. The separation of the terms having different algebraic
structure into different orders of the asymptotic expansion in $\afp$ is not  due to the
inherent properties of the Weyl tensor itself, but rather due to the asymptotic degeneracy of the
tetrad with respect to which the Weyl tensor is evaluated.
The coefficients in \eqref{wrongpeeling} are calculated in the \emph{interpretation tetrad}
which is parallelly transported along the null geodesic. We have seen that such a tetrad
becomes infinitely boosted with respect to a regular tetrad defined in terms
of the conformal geometry (see, e.g., relations \eqref{NullRotAuxExpl},
\eqref{OmegaInAfp}). The Weyl tensor evaluated in the tetrad which is defined using
the conformal techniques (i.e., the field calculated in the conformal geometry
and then appropriately rescaled to obtain the physical quantity) has a typical behaviour
${\WTP{}{}\sim\afp^{-3}}$ (cf.\ equation \eqref{fieldsnearscri}) and it does \emph{not}
exhibit any peeling-off behaviour. It is the transformation to the
interpretation tetrad (by the infinite boost, see equation \eqref{IntRelParField})
which gives rise to peeling-off of the components with a
different algebraic structure.

The field thus becomes asymptotically of type N only when viewed from the
parallelly transported tetrad, with the algebraically special direction oriented along
the tangent to the null geodesic approaching infinity. Already this
dependence of the algebraically special direction, along which the field asymptotically aligns, on the
direction of the geodesic indicates that the asymptotic algebraic degeneracy suggested by
\eqref{wrongpeeling} is not an invariant property of the field but an effect resulting
from specific relation between the field and the observer.

As we said, near a \emph{null} conformal infinity the magnitude of
leading coefficient ${\sim\afp^{-1}}$ in the
expansion \eqref{wrongpeeling} actually does \emph{not} depend on the direction of the
null geodesic (see \eqref{FieldInterpCompleteSum} for ${\nsgn=0}$), and can thus
be assigned a more invariant meaning --- we may speak about
non-radiative fields if this leading term is missing, and about radiative fields
otherwise. However, for a \emph{spacelike or timelike} conformal infinity
we have found that the magnitude of the leading term
\emph{does} depend substantially on the direction $\R$ of the geodesic.
Interestingly, such a dependence can be explicitly described in terms of the
principal null directions
of the field, see \eqref{FieldInterpCompleteSum} and the discussion in
sections~\ref{ssc:spacelikesri} and \ref{ssc:timelikescri}.

To summarize: the peeling-off behaviour of a field near a spacelike or timelike infinity
is not an invariant property of the field itself, but it is rather a statement
about the behaviour of the field components evaluated in suitable tetrads propagated parallelly
along null geodesics. For the full description of the components, the
standard \vague{peeling} needs to be supplemented by their directional dependence which
was presented above. We hope that our results may give some
clues to the understanding of radiation in spacetimes which are not asymptotically flat.

It is very difficult to obtain an explicit general relation between the matter
distribution and the corresponding distant gravitational field since the non-linearity of
the Einstein equations effectively allows gravitation to act as its own source. Therefore,
it  remains an open  problem to relate the structure of  bounded sources to
the principal null directions of the field at $\scri$ which essentially determines
the radiation structure at spacelike or timelike conformal infinities.
Some insight in this direction could hopefully be obtained by investigating
suitable exact model spacetimes.

\section*{Acknowledgments}

We are grateful to Ji\v r\'\i\  Bi\v c\'ak, who brought our attention to
the problem of radiation under the presence of a cosmological constant,
for many valuable comments and suggestions.
We also thank Jerry Griffiths and Marcello Ortaggio for careful reading of the manuscript.
This work was supported by the grant GA\v{C}R 202/02/0735 of the Czech Republic.

\appendix

\section{Asymptotic polyhomogenous expansions}
\label{apx:expansions}

In section~\ref{ssc:geodesics} we integrated the equation \eqref{RelatAfpCafp}
for a physical affine parameter, and we obtained its expansion \eqref{AfpOfCafp}
in terms of the conformal affine parameter ${\cafp}$. This can easily be inverted only
in the leading order, ${\cafp=-1/\afp}$. Here we derive the expansion
of conformal affine parameter ${\cafp}$ in terms of ${\afp}$ up to a higher order.

First, assuming smoothness of the conformal factor in conformal affine parameter
near $\scri$, we have (cf.\ equation \eqref{OmegaAnalyt})
\begin{equation}\label{apx:OmegaAnalyt}
\om=-\EPS\,\cafp+\om_2\,\cafp^2+\om_3\,\cafp^3+\ldots \commae
\end{equation}
where $\om_i$ are constants. Expanding ${\Omega^{-2}}$, the integration of \eqref{RelatAfpCafp} then leads to
\begin{equation}\label{apx:AfpOfCafp}
\afp=-\frac1{\cafp}
     +\bigl(2\EPS\, \om_2\,\ln\abs{\cafp}+\afp_0\bigr)
     +\bigl(3\om_2^2+2\EPS\om_3\bigr)\;\cafp
     +\ldots
\end{equation}
(cf.\ equation \eqref{AfpOfCafp}),
where ${\afp_0}$ is a constant of integration.
This expression contains the logarithmic term ${\,\ln\abs{\cafp}\,}$
which means that the relation between ${\afp}$ and ${\cafp}$ is intrinsically
non-analytic and cannot thus be inverted as a standard power expansion.
We have to look for an inverse expansion in a broader class of functions, namely
we admit functions which for small ${\xi}$ can be written as
\begin{equation}\label{PolyHom}
f(\xi) = \sum_{j=j_*}^\infty f_{j}(\ln^{\!-\!1}\!\abs{\xi})\;\xi^j\comma
\end{equation}
where ${j,j_*\in\integern}$, and the \vague{coefficient} ${f_{j}(\ln^{\!-\!1}\!\abs{\xi})}$
in the power expansion is
(infinite) polynomial of the reciprocal logarithm ${x=\ln^{\!-\!1}\!\abs{\xi}}$.
More precisely, ${f_{j}(x)}$ is a function which is analytic
(with a possible pole of a finite order ${-k_*}$ if ${k_*<0}$) at ${x=0}$,
\begin{equation}\label{PolyHomCoef}
f_{j}(x) = \sum_{k=k_*}^\infty f_{j,k}\;x^k\period
\end{equation}
Inspired by \cite{Chruscieletal:1995}, we may call such an expansions polyhomogenous.
The expansion \eqref{PolyHom} with the leading coefficient ${f_{j_*}(x)}$ regular and non-vanishing at ${x=0}$
can be substituted into another polyhomogenous expansion, and
the result remains again in the class of polyhomogenous expansions.

The expansion \eqref{apx:AfpOfCafp} is exactly of the form \eqref{PolyHom}
for a small parameter ${\cafp}$. We can seek the inverse relation as a polyhomogenous
expansion in the \emph{small} parameter ${\iafp=-\afp^{-1}}$
(i.e., in the reciprocal physical affine parameter; notice the difference between ${\iafp}$ and ${\EPS}$):
\begin{equation}\label{CafpOfAfpPolyHom}
\cafp=\iafp
      +\cafp_{2}(\ln^{\!-\!1}\!\abs{\iafp})\;\iafp^2
      +\cafp_{3}(\ln^{\!-\!1}\!\abs{\iafp})\;\iafp^3
      +\ldots \period
\end{equation}
Substituting into \eqref{apx:AfpOfCafp}, expanding logarithmic terms, and requiring that
the resulting expansion should lead to the single term ${\afp=-\iafp^{-1}}$, we find
\begin{align}
&\mspace{-34mu}\cafp=\iafp
      -\bigl(2\EPS\om_2\ln\abs{\iafp}+\afp_0\bigr)\;\iafp^2
\label{apx:CafpOfAfp}\\
& \mspace{-3mu}
      +\bigl(4\om_2^2\,\ln^2\!\abs{\iafp}+4\om_2(\EPS\afp_0+\om_2)\ln\abs{\iafp}
             +\afp_0^2+2\EPS\afp_0\om_2-3\om_2^2-2\EPS\om_3\bigr)\;\iafp^3
      +\ldots \period\notag
\end{align}
Thus, the conformal factor \eqref{apx:OmegaAnalyt} is
\begin{align}
&\mspace{-34mu}\om=-\EPS\,\iafp
      +\bigl(2\om_2\ln\abs{\iafp}+\EPS\afp_0+\om_2\bigr)\;\iafp^2
\label{apx:OmegaOfAfp}\\
& \mspace{-3mu}
      -\EPS\bigl(4\om_2^2\,\ln^2\!\abs{\iafp}+4\om_2^2(\EPS\afp_0+2\om_2)\ln\abs{\iafp}
+\afp_0^2+4\EPS\afp_0\om_2-3\om_2^2-3\EPS\om_3\bigr)\;\iafp^3
      +\ldots \period\notag
\end{align}

Integrating now equation \eqref{RelConfInterp} for  parameter ${\intTL}$,
in which we expand ${\om}$ and the right-hand side in parameter ${\cafp}$, see
equations \eqref{apx:OmegaAnalyt} and \eqref{Mepansion}, we obtain
\begin{equation}\label{apx:LinCafp}
\begin{split}
   \intTL &= M_{1}\ln|\cafp|+\intTL_{0} +(M_{2}+2\EPS M_{1} \om_2 )\, \cafp \\
   &\quad + {\textstyle\frac12}\,\bigl(M_3+2\EPS M_2 \om_2 + M_1 (3\om_2^2+2\EPS\om_3)\bigr)\,\cafp^2 + \dots \commae
\end{split}
\end{equation}
and expressing this in terms of the reciprocal physical affine parameter using \eqref{apx:CafpOfAfp}
\begin{equation}\label{apx:PhiLinAfp}
\begin{split}
   \intTL &= M_{1}\ln|\iafp|+\intTL_{0}
   + (-2 \EPS M_{1} \om_{2} \ln|\iafp| + M_{2} - M_{1} \afp_0 + 2 \EPS M_{1} \om_{2})\,\iafp\\
   &\quad+ \Bigl( 2 M_1\om_2^2\,\ln^2|\iafp| -2\EPS(M_2- M_1 \afp_0)\om_2\,\ln|\iafp|\\
   &\qquad+{\textstyle\frac12}\bigl(M_3-2M_2(\afp_0-\EPS\om_2)+M_1(\afp_0^2-3\om_2^2-2\EPS\om_3)\bigr)\Bigr)\,\iafp^2 + \ldots \period
\end{split}
\end{equation}
Moreover, the coefficients ${\om_i}$, ${M_i}$ in the expansions \eqref{OmegaAnalyt} and \eqref{Mepansion} can be
expressed in terms of derivatives of ${\om}$ and ${\,\bA^a\grad_a\om\,}$ with respect of
${\cafp}$. Namely, ${\om_2}$ and ${M_1}$ are given by
\begin{align}
\om_2&=\frac12 \frac{d^2\om}{\,d\cafp^2}\Big|_{\cafp=0}
     =\scale^2 \Bigl(\kP^b\kP^a \,\ccovd_b\grad_a\om \Bigr)\Big|_{\scri}   \commae\label{om2explicite}\\
M_1&={\sqrt2\scale}\,\frac{d}{d\cafp}\!\bigl(\mP^a \,\grad_a\om \bigr)\Big|_{\cafp=0}
   =2\scale^2\,\Bigl(\kP^b\mP^a \,\ccovd_b\grad_a\om \Bigr)\Big|_{\scri}   \commae\label{m1explicite}
\end{align}
where we used \eqref{ConfK} and \eqref{ConfParallel}. Employing equations
\eqref{RicciRelation} and \eqref{ConfscalarcurvRelation} we obtain
\begin{equation}\label{CovGrad}
\ccovd_b\grad_a\om  =\textstyle\frac{1}{4}
   \cmtrc_{ab}\cdalamb\om
   +\frac{1}{2}\,\om\,\bigl[(\Ric_{ab}-\frac{1}{4}\scR\,\mtrc_{ab})
     -(\cRic_{ab}-\frac{1}{4}\cscR\,\cmtrc_{ab})\bigr]   \period
\end{equation}
Consequently,
\begin{align}
\om_2&=\textstyle
    \frac{1}{2}\,\scale^2\,\left(\om\,\,\kP^b\kP^a \,
    \bigl[ (\Ric_{ab}-\frac{1}{4}\scR\,\mtrc_{ab})
    -(\cRic_{ab}-\frac{1}{4}\cscR\,\cmtrc_{ab})\bigr]
\right)\Big|_{\scri}   \commae\label{om2explicite2}\\
M_1&=\textstyle
    \scale^2\,\left(\om\,\,\kP^b\mP^a \,
    \bigl[ (\Ric_{ab}-\frac{1}{4}\scR\,\mtrc_{ab})
    -(\cRic_{ab}-\frac{1}{4}\cscR\,\cmtrc_{ab})\bigr]
\right)\Big|_{\scri}   \commae\label{LogTermMissing2}
\end{align}
We assume  regularity of the
conformal geometry near the infinity so that the second terms in brackets,
${\om(\cRic_{ab}-\frac{1}{4}\cscR\,\cmtrc_{ab})}$, vanish on $\scri$.
The first terms can be expressed as the specific tetrad components
of the traceless Ricci tensor \cite{Stephanietal:book}, namely
\begin{equation}\label{RicciComp}
{\RicciP{\auxT}{00}=\textstyle{\frac{1}{2}}(\Ric_{ab}-\textstyle{\frac{1}{4}}\scR\,\mtrc_{ab})\,\kG^b\kG^a}\comma\quad
{\RicciP{\auxT}{01}=\textstyle{\frac{1}{2}}(\Ric_{ab}-\textstyle{\frac{1}{4}}\scR\,\mtrc_{ab})\,\kG^b\mG^a}\commae
\end{equation}
which in view of Einstein equations \eqref{EinstEq} are proportional to the corresponding
components of the energy-momentum tensor.  We thus obtain
\begin{align}
\om_2&=\scale^2\,\left(\om^{-1}\RicciP{\auxT}{00} \right)\big|_{\scri}
   \sim \left(\cafp^{-1}\,\RicciP{\auxT}{00} \right)\big|_{\cafp=0}   \commae\label{om2explicite3}\\
M_1&=2\scale^2\,\left(\om^{-1}\RicciP{\auxT}{01} \right)\big|_{\scri}
   \sim \left(\cafp^{-1}\,\RicciP{\auxT}{01} \right)\big|_{\cafp=0}   \commae\label{LogTermMissing3}
\end{align}
where ${\RicciP{\auxT}{00}}$ and ${\RicciP{\auxT}{01}}$ are evaluated with respect to the tetrad \eqref{Auxiliary}.
These vanish identically for vacuum spacetimes. Moreover, ${\om_2}$ and ${M_1}$ are zero also in non-vacuum cases
such that near the conformal infinity the matter field decays  faster than ${\sim\cafp}$.
It corresponds to the situation when the Penrose's asymptotic Einstein condition
(equation \eqref{AsymEinstCond}, cf.\ (9.6.21) of \cite{PenroseRindler:book}) is satisfied.
With ${\om_2=0}$, ${M_1=0}$ the logarithmic terms in expansions \eqref{apx:CafpOfAfp}--\eqref{apx:PhiLinAfp}
disappear.

\section{Tetrads and fields in spinor formalism}
\label{apx:spinors}

Following, e.g., \cite{PenroseRindler:book}, the field of any spin
${s=0,\,\frac12,\,1,\,\frac32,\,\dots}$ can be represented using two-component symmetric
spinor ${\field}$ with ${2s}$ lower indices. To fix conventions for various signs and
prefactors which alter in the literature we first summarize some general relations for
spinors and their relation to tangent vectors.

Two-component spinors at a point ${x}$ form two mutually conjugated complex
vector spaces ${\SRspc_x\mfld}$ and ${\SLspc_x\mfld}$ of dimension two. We use capital
latin letters for indices of spinors from ${\SRspc_x\mfld}$ and letters with a bar
for the conjugated spinors. Spinor spaces are equipped with \defterm{skew-symmetric
metrics} ${\SRDe AB}$ and ${\SLDe AB}$ respectively, and with their inverses ${\SRUe AB}$ and
${\SLUe AB}$ (such that, e.g., ${\SRUe AM \SRDe BM=\SRid^{\SX A}_{\SX B}}$). These metrics
are used for lowering and raising indices:
$\tens{\psi}^{\SX A} = \SRUe AM\, \tens{\psi}_{\SX M}$,
$\tens{\psi}_{\SX A} = \tens{\psi}^{\SX M}\, \SRDe MA$.
The space of real bi-spinors (i.e., spinors ${\tens{\alpha}^{\SX A\SC{A}}}$
such that ${\tens{\alpha}{}^{\SX
A\SC{A}}=\bar{\tens\alpha}{}^{\SX\SC{A}A}}$) with metric
${-\SRDe AB \SLDe AB}$ is isometric to the space of tangent vectors with metric
spacetime ${\mtrc_{ab}}$ through the \defterm{soldering form} ${\SUsf aAA}$.
Relation of both metrics is
\begin{equation}
  \mtrc^{ab} = -\SUsf aAA\,\SUsf bBB\, \SRUe AB\,\SLUe AB\comma\quad
  \SRDe AB\, \SLDe AB = -\mtrc_{ab}\,\SUsf aAA\, \SUsf bBB\period
\end{equation}

A spinor frame ${\SRo,\,\SRi}$ is called \defterm{normalized}
if it satisfies
\begin{equation}\label{spinorframenorm}
  \SRUe AB = \SRo^{\SX A}\, \SRi^{\SX B} - \SRi^{\SX A}\, \SRo^{\SX B}\comma\qquad\text{i.e.,}\quad
  \SRo^{\SX A}\, \SRi^{\SX B}\, \SRe_{\SX AB} = 1\period
\end{equation}
We can associate a normalized spinor frame ${\SRo,\,\SRi}$ with any null tetrad
${\kG,\,\lG,\,\mG,\,\bG}$ in the following way:
\begin{equation}
\begin{aligned}
  \kG^{a} &= \SUsf aAA\, \SRo^{\SX A}\, \SLo^{\SX\SC A}\comma\quad&
  \mG^{a} &= \SUsf aAA\, \SRo^{\SX A}\, \SLi^{\SX\SC A}\commae\\
  \lG^{a} &= \SUsf aAA\, \SRi^{\SX A}\, \SLi^{\SX\SC A}\comma\quad&
  \bG^{a} &= \SUsf aAA\, \SRi^{\SX A}\, \SLo^{\SX\SC A}\period
\end{aligned}
\end{equation}

Special Lorentz transformation \eqref{kfixed}, \eqref{lfixed}, and \eqref{boostrotation}
correspond to transformations of normalized spinor frame which leave
\eqref{spinorframenorm} unchanged. Namely, for null rotation with ${\kG}$ fixed
we have
\begin{equation}\label{kfixedspinor}
\begin{aligned}
\SRo &= \SRo_{\refT} \comma&
  \kG &=\kO \comma&
  \mG &=\mO+L\, \kO\commae\\
\SRi &= \SRi_{\refT}+\bar{L}\,\SRo_{\refT} \comma&
  \lG &= \lO +\bar{L}\,\mO + L\,\mG +L\bar{L}\,\kG \comma&
  \bG &=\bO+\bar{L}\,\kO\commae
\end{aligned}
\end{equation}
and for null rotation with ${\lG}$ fixed,
\begin{equation}\label{lfixedspinor}
\begin{aligned}
\SRo &= \SRo_{\refT}+K\SRi_{\refT} \comma&
  \kG &= \kO +\bar{K}\,\mO + K\,\mG +K\bar{K}\,\kG \comma&
  \mG &=\mO+K\,\kO\commae\\
\SRi &= \SRi_{\refT} \comma&
  \lG &=\lO \comma&
  \bG &=\bO+\bar{K}\,\kO\period
\end{aligned}
\end{equation}
Boost and rotation are
\begin{equation}\label{boostrotspinor}
\begin{aligned}
\SRo &= B^{\frac12} \exp\bigl(i\frac{\intTphi}{2}\bigr)\,\SRo_{\refT}\comma&
  \kG &= B\, \kO\comma&
  \mG &=\exp\bigl(i\intTphi\bigr) \mO\commae\\
\SRi &= B^{-\!\frac12}\, \exp\bigl(-i\frac{\intTphi}{2}\bigr)\,\SRi_{\refT}\comma&
  \lG &= B^{-1} \lO\comma&
  \bG &=\exp\bigl(-i\intTphi\bigr) \bO\period
\end{aligned}
\end{equation}

As we have said, the field of spin ${s}$ can be represented by a spinor
${\field_{\SX A_1\cdots A_{2s}}}$ which is symmetric in all indices.
The space of such symmetric spinors
forms a representation space for the irreducible representation of type ${(0,s)}$
of the ${\mathsf{SL}(2\complexn)}$ group, or of its ${\mathsf{sl}(2\complexn)}$
Lie algebra which is isomorphic to Lie algebra of Lorenz group
${\mathsf{so}(1,3)}$.

Field equations for the zero-rest-mass field of spin ${s}$ are usually written in the form%
\begin{equation}
\SRUe MN\, \covd_{\SX M\SC A} \field_{\SX NA_2\cdots A_{2s}} = 0\comma\quad
\text{with}\quad \covd_{\SX A\SC A} = \SUsf aAA\,\covd_{a}\period
\end{equation}
It is well-known \cite{PenroseRindler:book} that such an equation is not
consistent for ${s>2}$ in a general curved background, and there are restrictions
on curvature to achieve consistency for ${s>1}$. However, the exact form of
the field equations is not necessary for our discussion. We only assume that we
may obtain the field ${\field}$ from some unspecified theory which prescribes
the behaviour of the field.

The examples are spinors ${\WTSR_{\SX ABCD}}$ and ${\EMSR_{\SX AB}}$ of
spin ${2}$ and ${1}$ which represent the gravitational and electromagnetic fields, respectively.
These spinors are related to the Weyl tensor ${\WT_{abcd}}$  as
\begin{equation}\label{Wspinor}
\begin{aligned}
  &\WT_{abcd} =  \SDsf aAA\, \SDsf bBB\, \SDsf cCC\, \SDsf dDD\,
    \bigl(\WTSR_{\SX ABCD}\,\SLDe AB\,\SLDe CD + \WTSL_{\SX\SC A\SC B\SC C\SC D}\,\SRDe AB\,\SRDe CD\bigr)\commae\\
  &\WTSR_{\SX ABCD} = {\textstyle\frac14}\,\SUsf aAA\, \SUsf bBB\, \SUsf cCC\, \SUsf dDD\,
    \WT_{abcd}\,\SLUe AB\, \SLUe CD\commae
\end{aligned}
\end{equation}
and to the electromagnetic tensor ${\EMF_{ab}}$ as
\begin{equation}\label{EMspinor}
\begin{aligned}
  &\EMF_{ab} = \SDsf aAA\, \SDsf bBB\,
    \bigl(\EMSR_{\SX AB}\,\SLDe AB + \EMSL_{\SX\SC A\SC B}\,\SRDe AB\bigr)\comma\\
  &\EMSR_{\SX AB}={\textstyle\frac12}\, \SUsf aAA\, \SUsf bBB\,\EMF_{ab}\,\SLDe AB\period
\end{aligned}
\end{equation}

The field ${\field}$ has ${2s+1}$ independent components. In the normalized spinor frame
${\SRo,\,\SRi}$ these can be identified as
\begin{equation} \label{spincompon}
  \fieldP{}{j} = \field_{\SX A_1\cdots A_j A_{j+1} \cdots A_{2s}} \,
  \SRi^{\SX A_1}\dots \SRi^{\SX A_j}\, \SRo^{\SX A_{j+1}}\dots \SRo^{\SX A_{2s}}\comma\quad
  j=0,\,1,\,\dots,\,2s\period
\end{equation}
Substituting the transformations \eqref{kfixedspinor}, \eqref{lfixedspinor}, and
\eqref{boostrotspinor} of the spinor frames into \eqref{spincompon} we immediately obtain
the transformation properties of the field components. Namely, we get
\begin{equation}\label{apx:kfixedField}
 \fieldP{}{j} = \fieldP{\refT}{j}+\binom{j}{1}\bar{L}\,\fieldP{\refT}{j-1}
  +\binom{j}{2}\bar{L}^2\,\fieldP{\refT}{j-2}+\binom{j}{3}\bar{L}^3\,\fieldP{\refT}{j-3}
  +\ldots+\bar{L}^j\,\fieldP{\refT}{0}
\end{equation}
for the null rotation with ${\kG}$ fixed, and
\begin{equation}\label{apx:lfixedField}
 \fieldP{}{j} = \fieldP{\refT}{j}+\binom{2s-j}{1}K\,\fieldP{\refT}{j+1}
  +\binom{2s-j}{2}K^2\,\fieldP{\refT}{j+2}+\ldots+K^{2s-j}\,\fieldP{\refT}{2s}
\end{equation}
for the null rotation with ${\lG}$ fixed. Finally, for the boost and the rotation we obtain
\begin{equation}\label{apx:boostrotationField}
 \fieldP{}{j} = B^{s-j}\,\exp\bigl(i(s-j)\intTphi\bigr)\; \fieldP{\refT}{j} \period
\end{equation}

\section*{References}
\label{sc:references}


\end{document}